\documentclass[sn-mathphys]{sn-jnl}
\jyear{2022}%

\theoremstyle{thmstyleone}%
\theoremstyle{thmstyletwo}%

\theoremstyle{thmstylethree}%

\usepackage{graphicx}
\usepackage{dcolumn}
\usepackage{bm}
\usepackage{siunitx}  
\usepackage{physics}
\usepackage{textcomp}
\usepackage{gensymb}
\usepackage{lineno}

\graphicspath{{Figures/}}

\begin{document}


\title[ ]{Time-resolved THz Stark spectroscopy}
\author[1]{\fnm{Bong Joo} \sur{Kang}}
\author[1]{\fnm{Egmont J.} \sur{Rohwer}}
\author[1]{\fnm{David} \sur{Rohrbach}}
\author[1]{\fnm{Maryam} \sur{Akbarimoosavi}}
\author[1]{\fnm{Zoltan} \sur{Ollmann}}
\author[1]{\fnm{Elnaz} \sur{Zyaee}}
\author[3]{\fnm{Raymond F.} \sur{Pauszek III}}
\author[4]{\fnm{Gleb} \sur{Sorohhov}}
\author[5]{\fnm{Alex} \sur{Borgoo}}
\author[5]{\fnm{Michele} \sur{Cascella}}
\author[1]{\fnm{Andrea} \sur{Cannizzo}}
\author[4]{\fnm{Silvio} \sur{Decurtins}}
\author[2]{\fnm{Robert J.} \sur{Stanley}}
\author[4]{\fnm{Shi-Xia} \sur{Liu}}
\author[1]{\fnm{Thomas} \sur{Feurer$^*$}}

\affil[1]{\orgdiv{Institute of Applied Physics}, \orgname{University of Bern}, \orgaddress{\city{Bern}, \postcode{3012}, \country{Switzerland}}}

\affil[2]{\orgdiv{Department of Chemistry}, \orgname{Temple University}, \orgaddress{\city{Philadelphia}, \postcode{19122}, \state{Pennsylvania}, \country{United States}}}

\affil[3]{\orgdiv{Department of Integrative Structural \& Computational Biology}, \orgname{The Scripps Research Institute}, \orgaddress{\city{La Jolla}, \postcode{92037}, \state{California}, \country{United States}}}

\affil[4]{\orgdiv{Department of Chemistry, Biochemistry and Pharmaceutical Sciences}, \orgname{University of Bern}, \orgaddress{\city{Bern}, \postcode{3012}, \country{Switzerland}}}

\affil[5]{\orgdiv{Department of Chemistry and Hylleraas Centre for Quantum Molecular Sciences}, \orgname{University of Oslo}, \orgaddress{\city{Oslo}, \postcode{N-0315 }, \country{Norway}}}

\affil[]{$^*$ corresponding author: thomas.feurer@unibe.ch}

\date{\today}
\abstract{
We introduce THz Stark spectroscopy by using intense single-cycle terahertz pulses as the electric field source and monitoring the induced spectral response of an isotropic molecular ensemble with a coincident femtosecond supercontinuum pulse. THz Stark spectroscopy offers several advantages over conventional Stark spectroscopy and opens previously inaccessible perspectives. Most importantly, THz pulses oscillate faster than typical molecular rotations and consequently eliminate the requirement to freeze the samples to prevent poling effects. Hence, THz Stark spectroscopy allows for time-resolved studies at arbitrary temperatures, specifically ambient conditions more relevant to physiological or operative conditions. Moreover, dynamical field effects, e.g., higher order Stark contributions or hysteresis effects (non-Markovian behavior), can be studied on the time scales of molecular vibrations or rotations. We demonstrate THz Stark spectroscopy for two judiciously selected molecular systems and compare the results to conventional Stark spectroscopy and first principle calculations.
}

\maketitle

\clearpage

\section{Introduction}

\begin{figure} [ht!]
\centering
\includegraphics[width=\columnwidth]{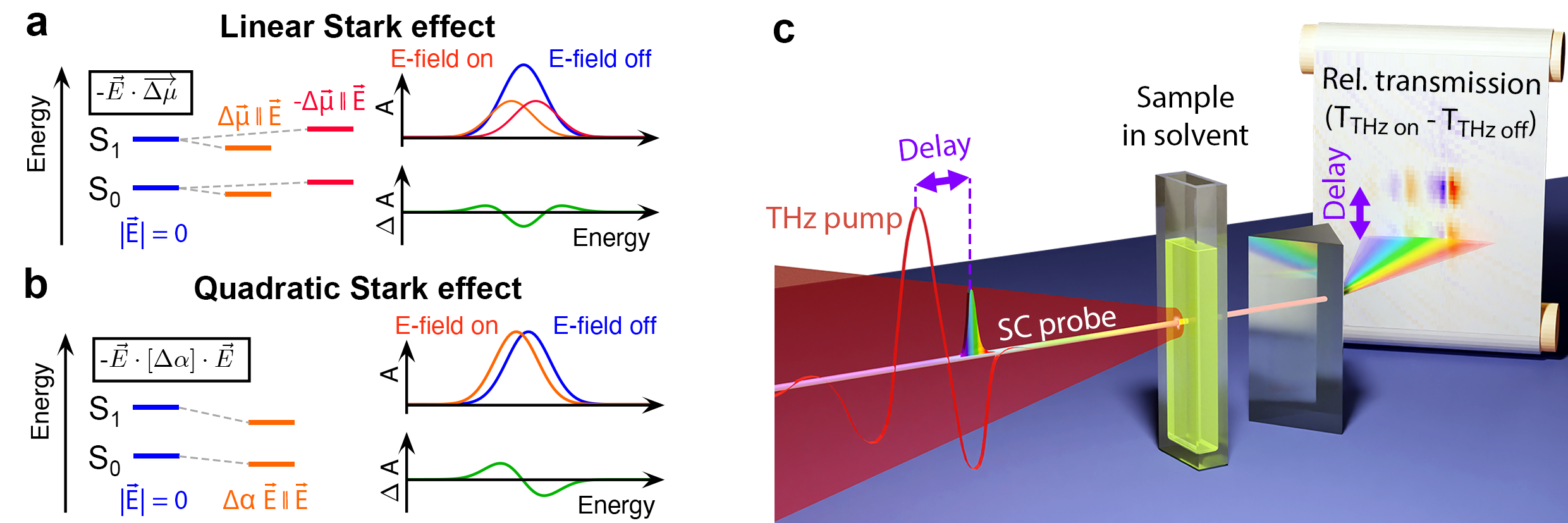}
\caption{\label{fig:SchematicPrinciple} \textbf{Electric field induced Stark effect and experimental concept.} Electric field induced Stark shift due to a difference in ground- and excited state dipole (\textbf{a}), i.e. linear Stark effect, and polarizability (\textbf{b}), i.e. quadratic Stark effect. For an isotropic distribution of molecules the ground state absorption band versus energy $A(E)$ broadens in the case of the linear Stark effect with the difference signal $\Delta A(E)$ resembling the second order derivative of the absorption band. In the case of the quadratic Stark effect the ground state absorption band shifts resulting in a difference signal $\Delta A(E)$ that is proportional to the first order derivative of absorption band. \textbf{c} Schematic representation of the experimental setup. The femtosecond supercontinuum probe pulse is scanned in time across the collinear single-cycle THz pulses and its spectrum is recorded by a spectrometer.}
\end{figure}

Stark spectroscopy is an invaluable tool to reveal information about physicochemical properties of molecules \cite{Brunschwig1998, Bublitz1997, Iimori2016, Karki1998, Liptay1969, Locknar1998, Mathies1976, Pauszek2013, Pein2019terahertz}. Sufficiently strong electric fields modify absorption spectra of isotropic ensembles of molecules if ground and excited state energy eigenvalues of the respective optical transition shift in energy due to the interaction with the applied electric field. In most cases, the interaction can be treated as a perturbation and expanded in a power series of the electric field. To first order, Stark spectroscopy reveals information on the electric dipoles, and to second order, on the induced dipoles, i.e. on the polarizabilities, of the two states. The former is referred to as the linear Stark effect (Fig.~\ref{fig:SchematicPrinciple}a), the latter is called the quadratic Stark effect (Fig.~\ref{fig:SchematicPrinciple}b). It is important to note that only a nonzero difference between ground and excited state dipole or polarizability results in a modified absorption spectrum, and consequently, only those differences can be extracted from a measurement. While a change in dipole moment, for instance, reflects the degree of charge separation or charge transfer associated with the transition, a change in polarizability describes the sensitivity of a transition to an external electric field \cite{Bublitz1997}. These effects are also known as electro-chromism \cite{Liptay1969} and a corresponding measurement gives insight in, for instance, photo-induced electron or charge transfer \cite{Hopkins2003, Oh1989, Reimers1991, Shin1995, Roiati2014stark}, nonlinear material properties \cite{Bublitz1997, Karki1998, Pauszek2013, Mehata2012}, biological organization and energy tuning \cite{Chowdhury2001, Kodali2009, Lockhart1988, Losche1987, Premvardhan2003, Somsen1998}, or solvato-chromism. Also, vibrational Stark spectroscopy of CO or CN ligands has developed into an important tool to measure in situ electric field strength in various chemical environments \cite{Verma2020}.

Conventional Stark spectroscopy uses low frequency (kHz) electric fields, which oscillate much slower than typical rotation times of molecules in solution. Hence, molecules must be immobilized in order to avoid alignment of their dipoles along the applied electric field, otherwise this poling effect would result in an overwhelming increase of the overall absorption and obscure any Stark signatures. Typically, this is achieved by freezing the solvent, which limits the range of solvents that can be used since these need to form optical glasses to avoid scattering of probe light. Moreover, freezing prevents molecules to be characterized in their natural, liquid or physiological environment. Finally, the sample geometry, with the electrodes on the front- and backside of the sample cuvette, leads to a non-optimal geometry with the angle between the polarization of the Stark field and the probe being far from the ideal value of 0~deg (typically 50 to 60~deg) \cite{Bublitz1997a}.

Here, we demonstrate that increasing the frequency of the oscillating electric field to the terahertz (THz) regime removes all of the above mentioned constraints and disadvantages of conventional Stark spectroscopy. Since a number of years it became possible to generate phase stable single- or few-cycle THz pulses with sufficiently strong electric fields (up to or even exceeding $10^6$~V/m) to induce measurable Stark shifts as well as femtosecond supercontinuum (fs-SC) probe pulses to time-resolve those transient Stark signatures \cite{Keiber2016electro, Knorr2017phase}. A schematic of the experimental realization is shown in Fig.~\ref{fig:SchematicPrinciple}c. Scanning the time delay between probe pulse and the THz waveform allows for the measurement of Stark signatures for positive or negative electric fields and from zero field up to the peak field strength of the THz pulse (see Supplementary Information 3.1: Experimental setup). At THz frequencies the electric field oscillates faster than typical molecular rotation times, hence for the first time, time-resolved Stark spectroscopy of molecules is performed without freezing the sample. Consequently, virtually any solvent can be used and molecules can be studied in their natural chemical or biological environment. Moreover, a wider temperature range becomes accessible, especially temperatures above the freezing point of the solvent but also room temperature or higher. The THz frequency of the field source also entails a higher dielectric breakdown allowing for electric field strengths, which were impossible to attain previously \cite{Jones1995, Krasucki1966, Zhang2018segmented}. At such field strengths, for instance, transition polarizability or hyper-polarizability may play a more important role, particularly for weakly allowed transitions. Such higher-order electric field effects may require a new theoretical concept, especially when the applied electric field can no longer be treated as a perturbation to the system's Hamiltonian or when magnetic effects need to be considered. THz Stark spectroscopy offers additional minor advantages, i.e., the sample thickness can be substantially increased and is limited only by absorption or velocity matching between the THz and the probe pulse. Additionally, arbitrary angles between the electric field vector and the probe polarization can be used and the absence of electrodes avoids potential redox chemistry in the pristine sample.

For the present THz Stark spectroscopic study, we selected two molecules and for comparison we performed quantum chemical calculations as well as conventional Stark spectroscopy. One molecule consists of a strong electron donor tetrathiafulvalene and an electron acceptor benzothiadiazole, showing an energetically low-lying intramolecular charge transfer state with a substantial change in dipole moment. The other is an anthanthrene derivative tetrasubstituted with silyl-protected acetylene groups to extend its $\pi$-conjugation, leading to an intense and sharp absorption band with a large change in polarizability.

\section{Results and discussion}


\subsection{Dynamics of the Stark signature}

\begin{figure} [ht!]
\centering
\includegraphics[width=0.8\columnwidth]{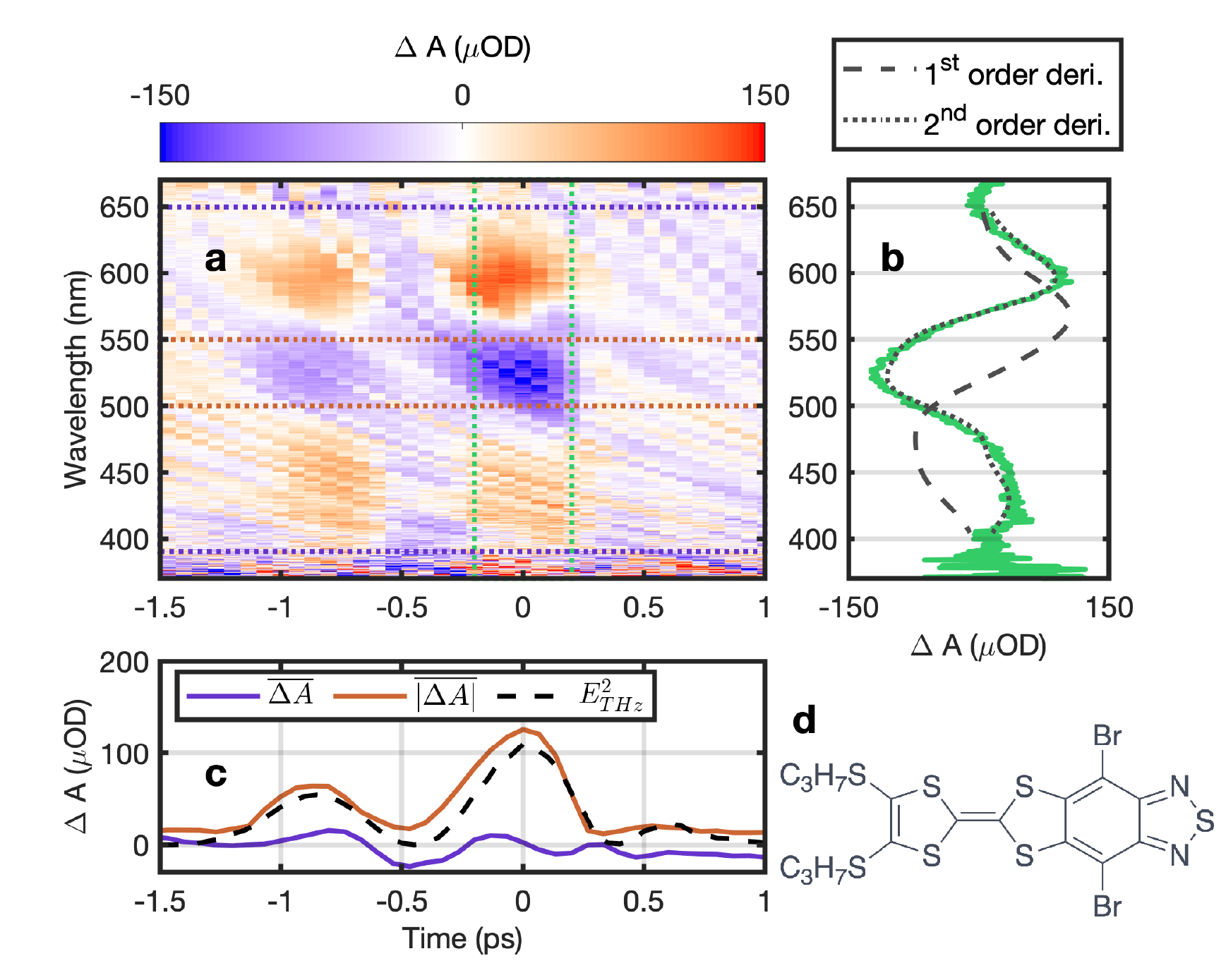}
\caption{\label{fig:TTF-BTD} \textbf{THz Stark signal of TTF-BTD with parallel orientation.} \textbf{a} False-color plot of the measured change in absorption spectrum as a function of time delay between THz and probe pulse and wavelength. \textbf{b} Time-averaged (between the two green dotted lines) change in absorption versus wavelength (green solid curve) compared to the scaled first (black dashed curve) and second order derivative (black dotted curve) of the ground state absorption spectrum. \textbf{c} Spectral average of the change in absorption between the purple dotted (purple curve) and red dotted lines (red curve) in \textbf{a}. The red curve is compared to the scaled square of the measured THz electric field $E_\mathrm{THz}^2$ (black dashed curve). \textbf{d} Chemical structure of TTF-BTD.}
\end{figure}

We first discuss the experimental results of a molecule with a pronounced intramolecular charge-transfer (ICT) character, namely an annulated electron donor-acceptor compound. Thereby, tetrathiafulvalene (TTF) acts as a strong donor and benzothiadiazole (BTD) equally as an acceptor within the compact and planar dyad \cite{Amacher2014, Pop2013a}. The bromine functionalization of BTD increases its acceptor strength and the propyl chains on TTF act as solubilizing groups (Fig.~\ref{fig:TTF-BTD}d) (see Supplementary Information 1.1: Synthesis and preparation). Figure~\ref{fig:TTF-BTD}a shows the color-coded THz-induced change in absorption $\Delta A(\tau,\lambda)$ in TTF-BTD versus time delay $\tau$ and wavelength $\lambda$ for parallel polarization orientation between probe pulse and THz waveform (corresponding results for a perpendicular orientation are provided in the Supplementary Information 3.4: Additional THz Stark spectroscopy results). The color scale indicates the difference in absorption in units of optical density (OD). The two-dimensional distribution reveals both temporal and spectral characteristics of the THz field-induced Stark shift. Time zero is set arbitrarily to the maximum change in absorption $\Delta A_\mathrm{max}$. The two peaks of the single-cycle THz pulse are well resolved with zero signal at the zero-crossing of the THz electric field. The measured spectra around the peaks of the THz field indicate a spectral broadening of the rather broad $S_0 - S_1$ absorption band between 390~nm to 650~nm, which is not unusual for ICT transitions. 

Calculating the margin along the delay axis between the two green dotted vertical lines results in the average Stark signal versus wavelength (green curve in Fig.~\ref{fig:TTF-BTD}b). Comparing the measured signal to the scaled first- and second-order derivative of the ground state absorption curve reveals that the Stark signature is dominated by a first order Stark contribution (black dotted curve) revealing a linear Stark effect caused by a change in dipole moment between ground and excited state. Additionally, we calculate two margins along the wavelength axis, first between the purple dotted horizontal lines resulting in the purple curve in (Fig.~\ref{fig:TTF-BTD}c), and second between the the two red dotted horizontal lines yielding the red curve in (Fig.~\ref{fig:TTF-BTD}c). The margin over the entire transition (purple curve) versus time delay is essentially zero confirming that no alignment of molecules to the applied THz field occurs. The residual small signal is most likely due to an imperfect correction of the group velocity dispersion in the fs-SC (see Supplementary Information 3.2: Group velocity dispersion correction) as well as a noise level of 20~$\mu$OD due to fs-SC fluctuations. These findings constitute an important result as they demonstrate that THz Stark spectroscopy can be applied to molecules in solution without the need to freeze the solvent. The red curve, averaged over one of the peaks in the Stark spectrum, indicates that the instantaneous Stark signal scales as the square of the THz field which is in agreement with Liptay's derivation of the linear and the quadratic Stark effect of an ensemble of molecules with isotropic orientation.

\begin{figure} [ht!]
\centering
\includegraphics[width=0.8\columnwidth]{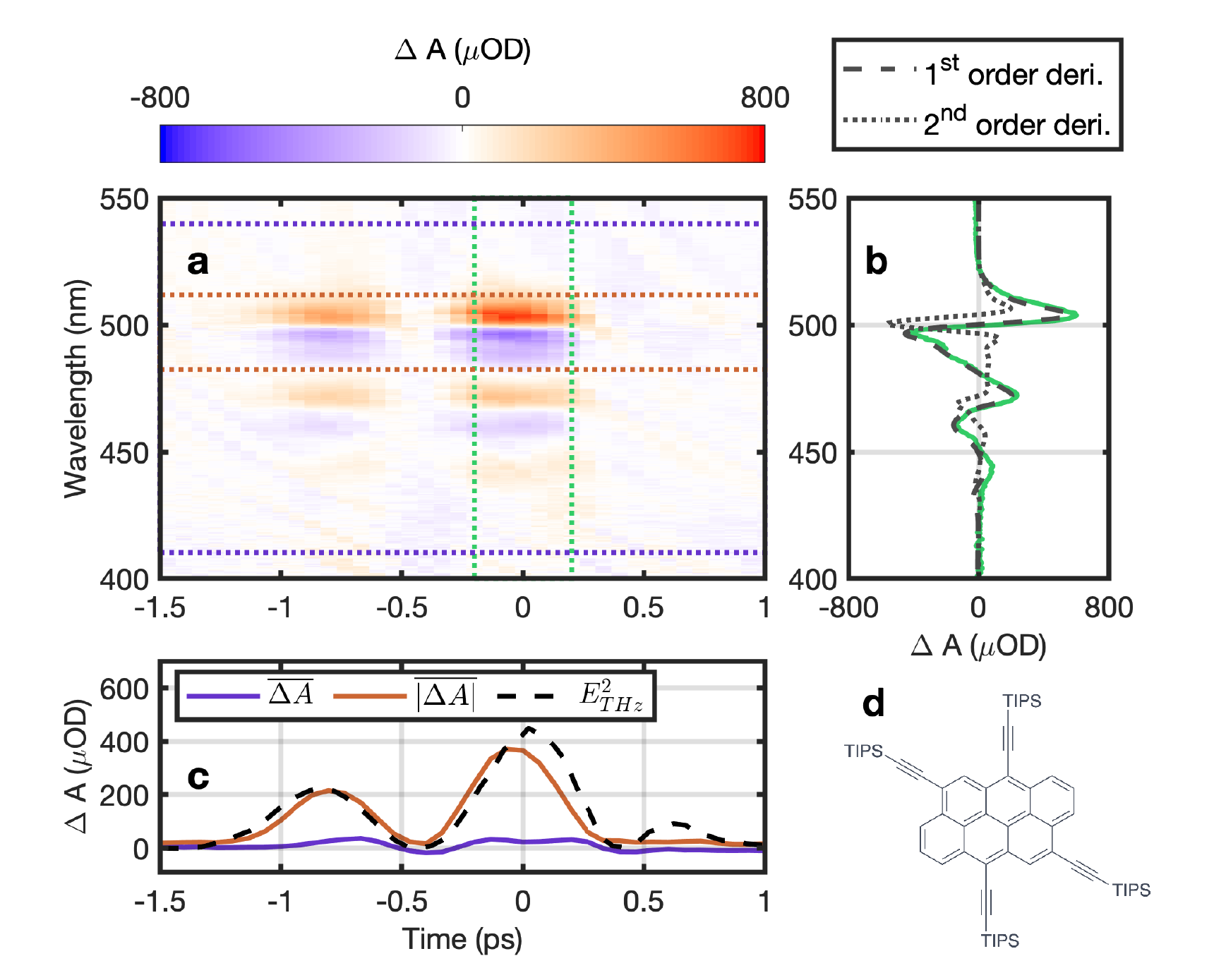}
\caption{\label{fig:Anthanthrene} \textbf{THz Stark signal of anthanthrene with parallel orientation.} \textbf{a} False-color plot of the measured change in absorption spectrum as a function of time delay between THz and probe pulse and wavelength. \textbf{b} Time-averaged (between the two green dotted lines) change in absorption versus wavelength (green solid curve) compared to the scaled first (black dashed curve) and second order derivative (black dotted curve) of the ground state absorption spectrum. \textbf{c} Spectral average of the change in absorption between the purple dotted (purple curve) and red dotted lines (red curve) in \textbf{a}. The red curve is compared to the scaled square of the measured THz electric field $E_\mathrm{THz}^2$ (black dashed curve). \textbf{d} Chemical structure of anthanthrene.}
\end{figure}

The second molecule, anthanthrene \cite{Giguere2013}, is a $\pi$-conjugated organic molecule of interest due to its semiconducting properties and potential applications in light emitting diodes or solar cells. The structure of anthanthrene is shown in Fig.~\ref{fig:Anthanthrene}d. It constitutes an interesting building block for organic electronics and Stark spectroscopy has the potential to reveal some of its relevant physicochemical properties. The transient Stark signal of the $S_0 - S_1$ transition of anthanthrene in toluene is shown in Fig.~\ref{fig:Anthanthrene}. The polarization between probe pulse and the THz waveform was parallel and the corresponding results for perpendicular polarization are provided in the Supplementary Information 3.4: Additional THz Stark spectroscopy results. The electronic transition shows a well separated vibrational progression, hence all vibrational bands are treated as one transition within our detection window. Margins are calculated following the same recipe as outlined above. A comparison of the time averaged Stark signature (Fig.~\ref{fig:Anthanthrene}b green curve) with the scaled first- (black dashed curve) and second-order derivative (black dotted curve) of the ground state absorption reveals a mostly quadratic Stark effect related to a difference in polarizability of ground and excited state. It also suggests that ground and excited state dipoles are, very likely based on symmetry arguments, negligible or similar in value. The two margins along the wavelength axis shown in Fig.~\ref{fig:Anthanthrene}c (purple and red curves) confirm that poling effects can be excluded, hence the sample maintains an isotropic distribution throughout, confirming again that the THz field oscillation is faster than any molecular rotation time.

That is, both molecules show a pronounced instantaneous Stark signature, which is well-matched to either the first- or the second-order derivative of the ground state absorption band, suggesting a dominant quadratic or linear Stark effect, respectively. In both cases, the Stark signal is proportional to the square of the THz electric field in agreement with Liptay formalism (see Supplementary Information 3.6: Liptay analysis) tracing the few-picosecond single-cycle THz waveform and we identify no measurable hysteresis or memory effect.

\subsection{Comparison between conventional and THz Stark spectroscopy}

For both molecules the Stark spectra for parallel and perpendicular polarization between probe pulse and THz waveform are fitted simultaneously and analyzed using the formalism outlined by Liptay \cite{Liptay1969, Rohwer2018} to extract relevant molecular parameters, such as differences in dipole moment or polarizability. We compare the results to those obtained from conventional Stark spectroscopy (see Supplementary Information 2: Conventional Stark spectroscopy) and those calculated from Density Functional Theory (DFT). DFT also provides molecular orbitals and associated energies for further discussion of results, specifically the charge redistribution associated with an excitation (see Supplementary Information 1.2: Density Functional Theory calculations). Recall that conventional Stark spectroscopy is done at 77~K, that is well below the freezing temperature of toluene, whereas THz Stark spectroscopy can be performed in principle at any temperature, but here is done at room temperature. To account for different optical path lengths, sample concentrations or electric field strength in the two measurement techniques, we compare the change in molar attenuation coefficient $\Delta\epsilon$ scaled to 1~MV/cm rather than the change in absorption. 

\begin{figure} [ht!]
\centering
\includegraphics[width=1\columnwidth]{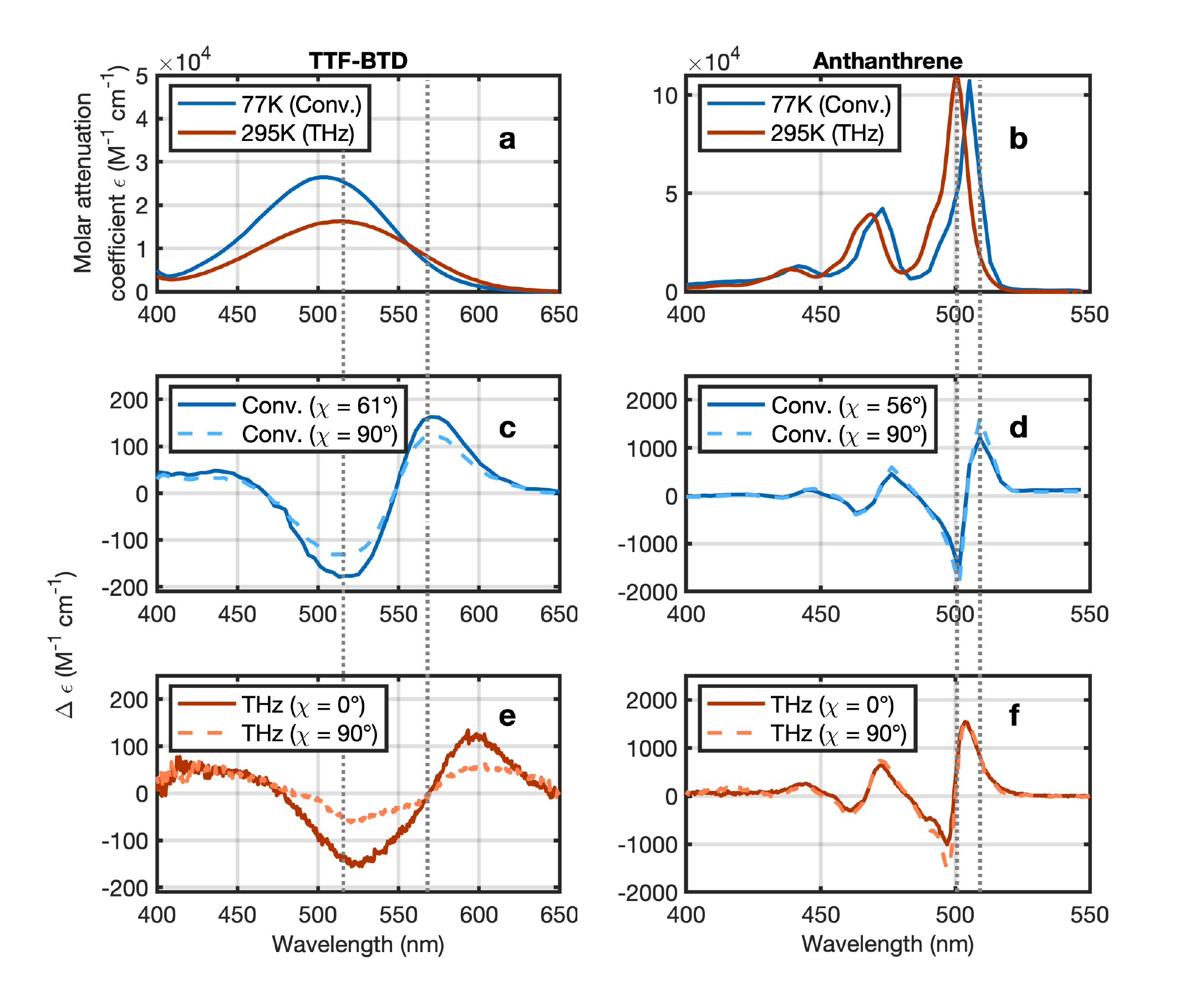}
\caption{\label{fig:SignalOverview_THzVsAC} \textbf{Comparison of conventional and THz Stark spectroscopy}. \textbf{a,b} Low temperature (77~K) and room temperature (295~K) absorption spectra of (a) TTF-BTD and (b) anthanthrene. \textbf{c,d} Conventional Stark spectra measured at 77~K for two values of the angle $\chi$ between the polarization of the THz and the probe pulses. \textbf{e,f} THz Stark spectra recorded at room temperature for parallel and perpendicular orientation of THz and probe polarization. For direct comparison the y-scale is in units of $\Delta \epsilon$ scaled to an electric field of 1~MV/cm. The grey dotted vertical lines are guides to the eye and help to visualize the shift of the spectra at the different temperatures.}
\end{figure}

The ground state absorption spectra of TTF-BTD and anthanthrene in toluene are shown in Fig.~\ref{fig:SignalOverview_THzVsAC}a and b for 77~K (blue curves) and room temperature (red curves). The DFT calculations suggest that TTF-BTD exhibits a broad absorption band due to a substantial ICT in the excited state. Anthanthrene, on the other hand, exhibits relatively narrow absorption features with an evident vibronic progression and DFT calculations mainly suggest a change in polarizability. When decreasing the temperature from 300~K to 77~K toluene is known to increase its effective polarity due to an increase in density \cite{Bublitz1998}. While in TTF-BTD this results in a blue-shift of the absorption peak from 514~nm to 505~nm due to the solvent's instantaneous electronic polarizability \cite{Akbarimoosavi2019}, in anthanthrene this leads to a increased stabilization of the energy levels and an associated red-shift of the lowest vibronic mode of the $S_0 - S_1$ HOMO - LUMO transition from 500~nm to 505~nm. Irrespective of molecule, thermal broadening of the absorption features is evident for increasing temperatures. Both effects illustrate the importance of fitting Stark spectra with ground state absorption spectra recorded at the same temperature.

Figure~\ref{fig:SignalOverview_THzVsAC}c and d show the change in molar attenuation coefficient $\Delta\epsilon$ scaled to 1~MV/cm for TTF-BTD and anthanthrene as measured by conventional Stark spectroscopy for a relative orientation between probe pulse and THz waveform polarization of 56~deg/61~deg (blue solid curve) and 90~deg (blue dashed curve). Note, that angles smaller than 56~deg are difficult to realize due to above mentioned geometrical constraints. The corresponding results of the THz Stark spectroscopy are shown in Fig.~\ref{fig:SignalOverview_THzVsAC}e and f. THz results were averaged over a 467~fs time window around the larger peak of the THz pulse. Unlike conventional Stark spectroscopy, the THz variant has no geometrical constraints and allows for angles down to 0~deg, which helps to improve the dynamic range of measurement as well as the accuracy of the Liptay analysis. Except for the temperature related blue- or red-shift and the different minimal angle, the Stark signatures of both methodologies are in excellent agreement with each other. The peak change in molar attenuation coefficient for TTF-BTD is slightly lower in the THz measurement due to the thermal broadening of the absorption spectrum at room temperature.

\subsection{Relevant molecular parameters}

A more rigorous comparison becomes available after analyzing the spectra in Fig.~\ref{fig:SignalOverview_THzVsAC} following the Liptay protocol. It reveals the change in dipole moment, $\Delta \mu$, the angle between the change in dipole moment and the transition dipole moment $m$, $\zeta$, the average change in polarizability, Tr($\Delta \alpha$), and the change in polarizability parallel to the transition dipole moment $m$, $m \; \Delta\alpha \; m$. The latter consistently showed large fitting errors and is not reported here. Table~\ref{tab:ExtractedParameter} summarizes these molecular parameters calculated from DFT as well as measured and extracted via the Liptay analysis of both Stark spectroscopy methodologies.

\begin{table}[ht!]
\caption{\label{tab:ExtractedParameter} Comparison of relevant molecular parameters as calculated via DFT or measured by conventional and THz Stark spectroscopy.}
 $ $
\begin{tabular}{ |p{2.5cm}||p{2.5cm}|p{2.5cm}|p{2.5cm}| }
\hline
Parameter & DFT & Conventional Stark & THz-Stark \\
\hline
\multicolumn{4}{|c|}{TTF-BTD: tetrathiafulvalene-benzothiadiazole} \\
\hline
$\Delta \mu$ (D)   & 16.2    & $14.7 \pm 0.1$ &   $15.3 \pm 1.8$ \\
$\zeta$ ($^\circ$)   & $\approx 0$    & $24.8 \pm 0.2$ &   $18.7 \pm 1.1$ \\
\hline
\hline
\multicolumn{4}{|c|}{anthanthrene: 4,6,10,12-tetrakis(triisopropylsilylethynyl)anthanthrene} \\
\hline
Tr($\Delta \alpha$) (\AA$^3$)   & 457    & $363 \pm 20$ &   $296 \pm 70$ \\
\hline
\end{tabular}
\end{table}

The agreement within the error bars in all molecular parameters extracted from the two Stark spectroscopy modalities is remarkable and confirms that both techniques deliver results which are in quantitative agreement with each other. The biggest source of error comes from the uncertainty of the electric field applied to the sample (see Supplementary Information 3.5: Characterization of THz pulses). Furthermore, the experimental results are in reasonable agreement with DFT calculations given the fact that the solvent environment can only be treated approximately if at all. Including the solvent effect in the DFT calculations for TTF-BTD leads to marginal modifications to the molecular geometry as compared to the gas phase and details can be found in reference \cite{Rohwer2018}. The mismatch in measured and calculated angle $\zeta$ most likely is due to these structural modifications, but may be also a result of temperature affecting electrostatic interactions and molecular geometry \cite{Rohwer2018, rondi2015solvation}. For the anthanthrene, the DFT calculations were performed at 0~K and without solvent effects, nevertheless, the experimental results agree very well with the DFT calculations. Hence, neither the low polarity of the solvent nor the increased temperature seem to drastically affect the dipole moment or the electronic polarizability of the molecule.

\section{Conclusion}

In conclusion, we have demonstrated that THz Stark spectroscopy indeed reveals the same physicochemical properties of molecules as conventional Stark spectroscopy, but at the same time opens hitherto inaccessible opportunities, because it is not subjected to the same limitations that apply to conventional Stark spectroscopy. Geometrical constraints are removed allowing for arbitrary angles between probe pulse and THz waveform polarization, no electrodes are required, which helps to avoid potential redox chemistry in the pristine sample, and the much higher frequency of THz waveforms allows for higher electric field strengths before the threshold for dielectric breakdown is reached. Most importantly however, THz Stark spectroscopy removes the need to immobilize the molecular ensemble by freezing the solvent. Hence, molecules or bio-molecules can now be studied in their natural environment and at relevant temperatures. Our findings are based on measurements of two molecules relevant in the context of molecular electronics. 

In principle, THz Stark spectroscopy will allow to observe transient or non-equilibrium electronic properties of molecules with sub-100~fs resolution. Consequently, THz Stark spectroscopy can be used to study molecular ensembles at conditions not accessible to conventional Stark spectroscopy, for instance, within a much increased range of temperatures or in different non-polar or polar solvents, even those that do not form transparent glasses at low temperatures. Today's high-field THz sources generate field strengths in excess of 1~MV/cm so that higher-order Stark contributions may become observable, such as non-Markovian responses (hysteresis effects) originating from electron‑phonon couplings. Higher order Stark contributions are impossible to access via conventional Stark spectroscopy but are relevant to model electron dynamics induced by external or local fields (e.g. charge and electron transfer) or to refine quantum chemistry codes. Moreover, the intrinsic time resolution of around 100~fs facilitates studies on the time-dependent physicochemical properties of a molecule during its photo-cycle, specifically it allows for Stark spectroscopy of excited states.

\section{Methods}
\textbf{THz Stark spectrometer}

The THz Stark spectrometer was designed to record the change in absorption $\Delta A(\tau,\lambda) = A_\text{THz on}(\tau,\lambda) - A_\text{THz off}(\tau,\lambda)$ as a function of time delay between the THz waveform and the probe pulse $\tau$ and of wavelength $\lambda$. The recorded Stark maps $\Delta A(\tau,\lambda)$ were subsequently corrected for the fs-SC group velocity dispersion (see Supplementary Information 3.2: Group velocity dispersion correction) and the background resulting from the pure solvent (see Supplementary Information 3.3: Background measurements). The analysis of the Stark spectra, here at maximum electric field, $\Delta A(\lambda)$, was outlined in reference \cite{Liptay1969}. After having identified the Stark-active transitions, the Stark spectra were subsequently analyzed with the Liptay formalism, which links the molar absorption $\Delta \epsilon(\bar{\nu})$ as a function of wavenumber to the ground state absorption spectrum $\epsilon(\bar{\nu})$. The model assumes a fixed angle between electric field and probe polarization and an isotropic distribution of transition dipole moments, which is achieved by freezing the sample in conventional Stark spectroscopy. The ground-state absorption spectra and two Stark spectra for different probe polarizations were fitted simultaneously with a weighted sum of the zeroth, first, and second order derivative of the ground state absorption spectrum. 

\begin{equation}
\Delta\epsilon(\bar{\nu}) = f_l^2 \abs{\vb{E}}^2 \left\{ a \epsilon(\bar{\nu})  + b \dv{\bar{\nu}} \left( \frac{\epsilon(\bar{\nu})}{\bar{\nu}} \right) + c \dv[2]{\bar{\nu}} \left( \frac{\epsilon(\bar{\nu})}{\bar{\nu}} \right) \right\}.
\end{equation}

From the fit parameters $a$, $b$, and $c$ we extracted the trace of the polarizability tensor, its projection along the transition dipole moment, the angle between the applied electric field and the probe polarization, the change in dipole moment, and the angle between the change in dipole moment and the transition dipole moment. An important ingredient to the fit is the THz electric field strength $\abs{\vb{E}}$ in the sample at which the probe pulse interrogates the molecular system. In order to account for all experimental effects we first measured the THz electric field in air and then performed finite difference time-domain simulations to determine the time dependence of the THz electric field experienced by the probe pulse in the complex cuvette/liquid environment. We found that the time dependence is almost identical, however with the peak electric field strength being reduced by a factor of 0.7 due to Fresnel reflections and Fabry-Perot effects. Hence, the maximum field in the sample was reduced to \SI[separate-uncertainty=true]{280 \pm 17}{\kV/\cm}. The electric field experienced by the molecules is further modified by the local field correction factor $f_l$, which is a measure of how the solvent cavity affects the field inside the cavity containing the molecule \cite{Aubret2019, Stanley2001, Premvardhan1998}. We estimated the local field correction factors for TTF-BTD and anthanthrene to 1.30 and 1.33 for conventional Stark spectroscopy and to 1.26 and 1.29 respectively for THz Stark spectroscopy (see Supplementary Information 3.7: Local field correction factor).


\section*{Declarations}

\subsection{Contributions}

M.A., E.J.R. and R.J.S. performed the conventional Stark experiments and suggested ways to improve the analysis. Z.O. built and characterized the high-field THz source. B.J.K., E.J.R., and D.R. performed the THz-Stark experiments and analyzed the data. G.S., S.D., and S.-X.L. synthesized the molecular samples. A.B. and M.C. performed the DFT calculations. All authors contributed to the discussion, writing and reviewing of the final manuscript.

\subsection{Corresponding author}
Correspondence to Thomas Feurer: thomas.feurer@unibe.ch

\subsection{Ethics declarations}
The authors declare no competing interests.

\subsection{Data availability}

The data supporting the findings of this study are available from the authors upon reasonable request.

\subsection{Acknowledgements}

We acknowledge experimental support by Steven E. Meckel. This work was supported by the National Center of Competence in Research - Molecular Ultrafast Science and Technology (NCCR MUST), a research instrument of the Swiss National Science Foundation as well as by the Swiss NSF (200021-204053). B.J.K. also acknowledges funding from the European Union’s Horizon 2020 research and innovation program under the Marie Skłodowska-Curie grant agreement (FP-RESOMUS - MSCA 801459). A.B. and M.C. acknowledge the support of the Research Council of Norway through the CoE Hylleraas Centre for Quantum Molecular Sciences (grant no. 262695), and the Norwegian Supercomputing Program NOTUR (grant no. NN4654K). 

\subsection{Authors and Affiliations}

\begin{enumerate}
   
\item Institute of Applied Physics, University of Bern, 3012 Bern, Switzerland

Bong Joo Kang, Egmont J. Rohwer, David Rohrbach, Maryam Akbarimoosavi, Zoltan Ollmann, Elnaz Zyaee, Andrea Cannizzo \& Thomas Feurer

\item Department of Chemistry, Temple University, Philadelphia, Pennsylvania 19122, United States
    
Robert J. Stanley

\item Department of Integrative Structural \& Computational Biology, The Scripps Research Institute, La Jolla, California 92037, United States
    
Raymond F. Pauszek III
    
\item	Department of Chemistry, Biochemistry and Pharmaceutical Sciences, University of Bern, 3012 Bern, Switzerland
    
Gleb Sorohhov, Silvio Decurtins \& Shi-Xia Liu

\item	Department of Chemistry and Hylleraas Centre for Quantum Molecular Sciences, University of Oslo, N-0315 Oslo, Norway
    
Alex Borgoo \& Michele Cascella

\end{enumerate}

\section[]{Supplementary information}

\subsection{Molecular systems}
\label{ssec:Samples}

\subsubsection{Synthesis and preparation}

The two molecular systems were selected because they show either a pronounced change in dipole moment, which allows us to unambiguously observe the linear Stark effect, or a pronounced change in polarizability, which allows us to observe the quadratic Stark effect. In addition, both molecules are relevant in the field of molecular electronics and the extracted physicochemical properties are relevant in their own right.

The first molecular system is fused heterocyclic tetrathiafulvalene \cite{Bendikov2004, Martin2013, Wu2009, Bergkamp2015, Segura2001} – benzothiadiazole \cite{JustinThomas2004, Wu2013, Belton2013} (TTF-BTD) with Br substituted at the 4 and 8 positions of BTD \cite{Pop2013}. This molecule is known to undergo intramolecular charge transfer \cite{Alemany2015, Geng2014} with a correspondingly large change in dipole moment upon excitation of the HOMO-LUMO transition. The relatively broad absorption band centered around 500~nm is a typical signature of such intramolecular charge transfer.

The second molecular system is the 4,6,10,12-tetrakis(triisopropylsilylethynyl)-anthanthrene compound \cite{Giguere2013}, which bears a graphene-like aromatic skeleton and is well-known for its quantum interference effect on the single molecule conductance, \cite{Lambert2018, Geng2015, Famili2019} and for its versatile chromophoric properties in organic electronics \cite{Geng2015a, Giguere2015, Zhang2012a, Shah2006, Shah2005}. It exhibits optically allowed transitions with significant oscillator strengths in the spectral range around 500~nm.

For all experiments, toluene was used as solvent because it has a low polarity and forms a transparent optical glass at 77~K, which is mandatory for the conventional Stark measurements. The low polarity limits effects due to varying solvent polarity at different temperatures and the potential for THz field-induced orientation of the solvent. For conventional Stark measurements, the sample concentration was 1~mM for TTF-BTD and 0.45~mM for anthanthrene. The sample concentration for the THz Stark measurements was 1~mM for TTF-BTD and 0.5~mM for anthanthrene. In the THz Stark experiment, a spectrosil quartz flow cell from Starna Scientific was used with a sample thickness of 200~$\mu$m and a wall thickness of 200~$\mu$m. Even though a flow cell is not specifically needed, it helped to suppress air bubble formation. Furthermore, any potential effects due to toluene evaporation are reduced. 

While the ground state absorption spectra in conventional Stark spectroscopy were measured directly with the Stark spectroscopy apparatus (with no applied electric field at low temperature), the room temperature ground state absorption spectra were recorded separately with a spectrophotometer (Perkin-Elmer Lambda 750 Spectrometer, 1~mm thick cuvette).

\subsubsection{Density Functional Theory calculations}

We performed DFT calculations of both systems to gain further insight into the charge redistribution between the molecular orbitals involved in the observed optical transitions and the associated energies. While details on our DFT computations on TTF-BTD have already been published elsewhere \cite{Rohwer2018}, we here present methods and results on the second system anthanthrene. 

\begin{figure} [ht!]
\centering
\includegraphics[width=0.7\columnwidth]{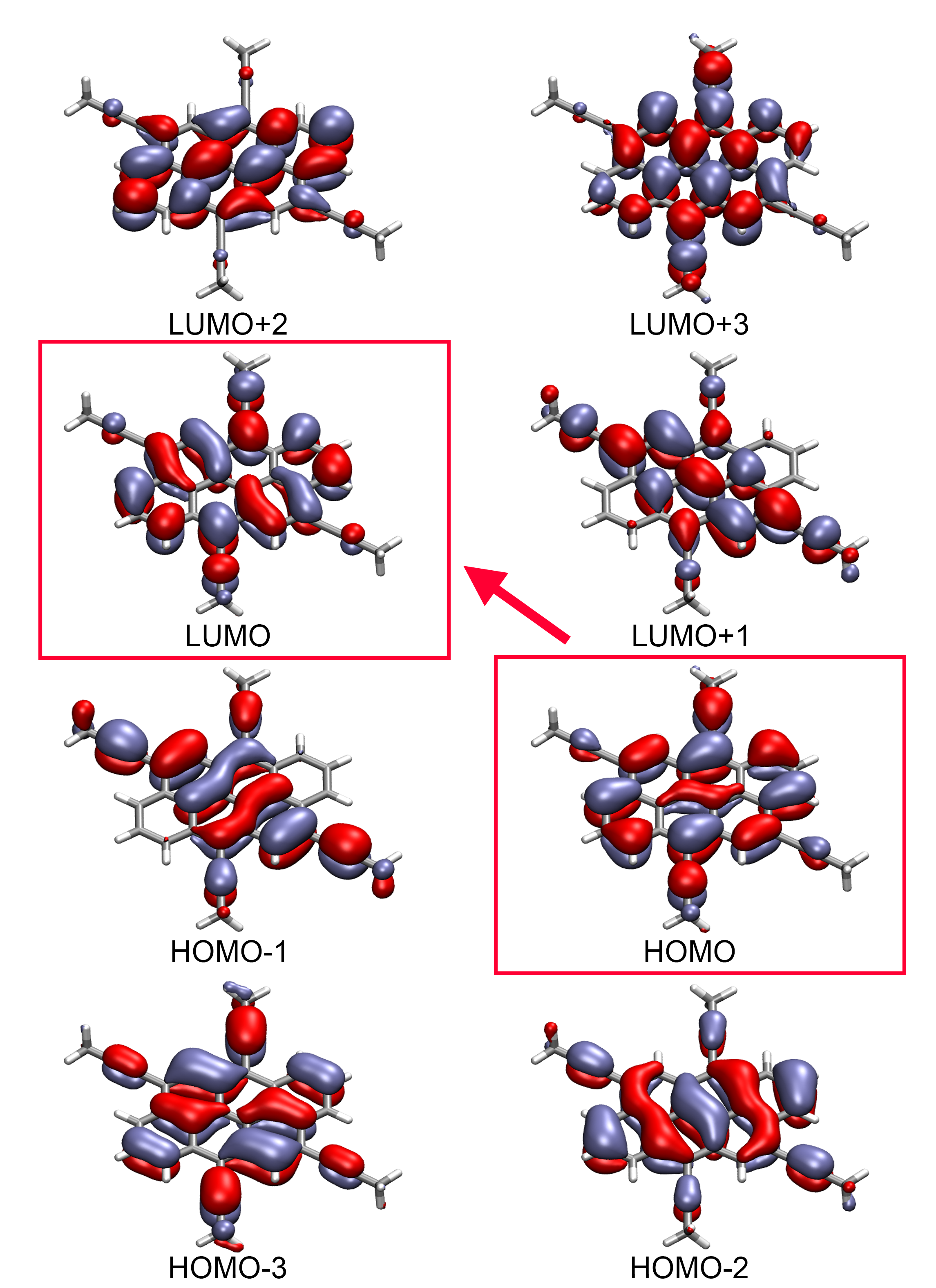}
\caption{\label{fig:DFT results} Molecular orbitals of anthanthrene. With reference to the main text, we draw the reader’s attention to the HOMO$\rightarrow$LUMO transition in particular.}
\end{figure}

The molecular geometries of anthanthrene were optimised at the Kohn-Sham DFT level. We used the PBE0 \cite{Adamo1999} or the B3LYP \cite{Becke1993} functionals to approximate the exchange-correlation energy in combination with the 6-31+G(d,p) basis set \cite{Hehre1972}. The excitation energies were computed by time-dependent DFT (TD-DFT) \cite{Runge1984}. The properties of excited states were obtained as higher-order response properties of the ground state. Specifically, the polarizability of an excited state was determined by first converging the electronic energy of the ground state and then by computing the double residue of the cubic response function as described in reference \cite{Jansik2004} and implemented in the DALTON software package \cite{Aidas2014}. Table~\ref{tab:DFT} summarizes possible transitions originating from the ground $S_0$ state with corresponding wavelength, oscillator strength, major molecular orbital contributions, and transition dipole moment. Molecular orbitals and associated energies involved in the calculated transitions for the sample are illustrated in Fig.~\ref{fig:DFT results}. 

\begin{table}
\caption{\label{tab:DFT} The ground state transitions of anthanthrene with wavelength, oscillator strength, and major contributing molecular orbitals calculated with TD-DFT.}

\begin{tabular}{ |p{1.4cm}||p{2cm}|p{2.3cm}|p{4.5cm}| }
 \hline
 Excited State	& Wavelength (nm) &	Oscillator strength & Major contributions (\%) \\
 \hline
 $\mathbf{S_1}  $ & \textbf{500.2} & \textbf{0.4846} & \textbf{HOMO} $\rightarrow$ \textbf{LUMO} \textbf{Pure} \\
 \hline
 $S_2$   & 419.7    & 0.0    &   HOMO-1$\rightarrow$LUMO 36 HOMO$\rightarrow$LUMO+1	~~~~~~~13 \\
 \hline
 $S_3$   & 402.8    & 0.0    &   HOMO-1$\rightarrow$LUMO 13 HOMO$\rightarrow$LUMO+1	~~~~~~~36 \\
 \hline
 $S_4$   & 395.2    & 0.0087 &   HOMO-2$\rightarrow$LUMO 27 HOMO$\rightarrow$LUMO+2	~~~~~~~22 \\
 \hline
 $S_5$   & 333.0    & 0.6133 &   HOMO-3$\rightarrow$LUMO 3   HOMO-2$\rightarrow$LUMO	~~~~~~~20   HOMO-1$\rightarrow$LUMO+1 2 HOMO$\rightarrow$LUMO+2	~~~~~~~22 \\
 \hline
 $S_6$   & 315.4    & 0.0366 &   HOMO-3$\rightarrow$LUMO 33   HOMO$\rightarrow$LUMO+2	~~~~~~~2   HOMO$\rightarrow$LUMO+3 ~~~~~~~12 \\
 \hline
\end{tabular}
\end{table}

The DFT calculations predict one optically allowed transitions with significant oscillator strength in the our spectral observation window, which is the $S_0 \rightarrow S_1$ transition. Due to the symmetry of the molecule and the symmetry of the molecular orbital distributions associated with this transition, we do not expect a permanent dipole in any of these states and as such we also do not expect a change in dipole moment over the transition. The computed polarizability changes are summarized in Table~\ref{tab:Deltaalpha}. For completeness we also include the results for the $S_0 \rightarrow S_5$ transition, which has an appreciable oscillator strength but is outside of the spectral observation window of the experiment.

\begin{table}
\caption{\label{tab:Deltaalpha} $\Delta\alpha$ computed by TD-DFT with an aug-cc-PVDZ basis set.}
\begin{tabular}{ |p{1.5cm}||p{3cm}|p{3cm}|p{3cm}|  }
 \hline
 State & Wavelength (PBE0)   & Wavelength (B3LYP)    & Tr($\Delta\alpha$) (B3LYP) \\
 \hline
 $S_{1}$   & 500.2~nm    & 518.8~nm &   457~${\si\AA}^{3}$ \\
 \hline
 $S_{5}$   & 333.0~nm    & 347.0~nm &   333~${\si\AA}^{3}$\\
 \hline
\end{tabular}
\end{table}

The average isotropic change in polarizability over the transitions was calculated using both PBE0 and B3LYP functionals. We checked convergence with the basis set (using aug-cc-PVTZ and daug-cc-PDVZ). We verified that using aug-cc-PVDZ the basis set error affects the quantitative result by less than 1\%. In addition, note that the lowest energy transition $S_0 \rightarrow S_1$ splits in a well-separated vibronic progression.
 
\subsection{Conventional Stark spectroscopy}

To understand the limitations of conventional field Stark spectroscopy we will describe the technique in general terms here. A more detailed description of the specific Stark spectrometer used, along with technical details can be found in a previous publication \cite{Rohwer2018}. Typically, a high voltage source is connected to transparent electrodes (indium tin oxide: ITO) on the inner front and back surface of the sample cell. The front and the back window are separated by a Kapton spacer with a thickness of \SI{25}{\um} and the sample cavity is filled by injection with the sample solution. The sample is then mounted in a cooled Dewar at 77~K to freeze the solvent. Freezing the sample is also a common work-around to increase the breakdown voltage and prevent unwanted redox chemistry. During the experiments, a 3.5~kHz sinusoidal signal is applied across a cell. While for TTF-BTD the applied voltage was $V_\mathrm{rms} = \SI{250}{\V}$ (with a peak field of 141~kV/cm), for anthanthrene the applied voltage was $V_\mathrm{rms} = \SI{150}{\V}$ (with a peak field of 85~kV/cm). A lamp and monochromator provide tunable probe light. The electric field-induced transmission changes are monitored at twice the AC frequency by a lock-in amplifier. The difference between the in- and out-of-phase components is plotted as a function of monochromator wavelength, resulting in a Stark spectrum. The cell can be rotated about a vertical axis within the Dewar to facilitate different relative angles between the applied field and the probe light polarization. The polarization of the probe light is set with a Glan-Taylor prism.

\subsection{THz Stark spectroscopy}

This section describes in detail all relevant steps to extract the molecular physicochemical constants. We start with a brief description of the experimental apparatus and the data recorded by it. Next, we outline how we correct for the group velocity dispersion of the probe pulses and give a detailed account of the background subtraction protocol. A further important ingredient for the analysis is the THz electric field strength in the sample and we determine it by a combination of measurements and finite difference time domain simulations. Finally, the measurements are analyzed by the Liptay formalism, which is briefly described toward the end. 

\subsubsection{Experimental setup}

\begin{figure} [ht!]
\centering
\includegraphics[width=0.8\columnwidth]{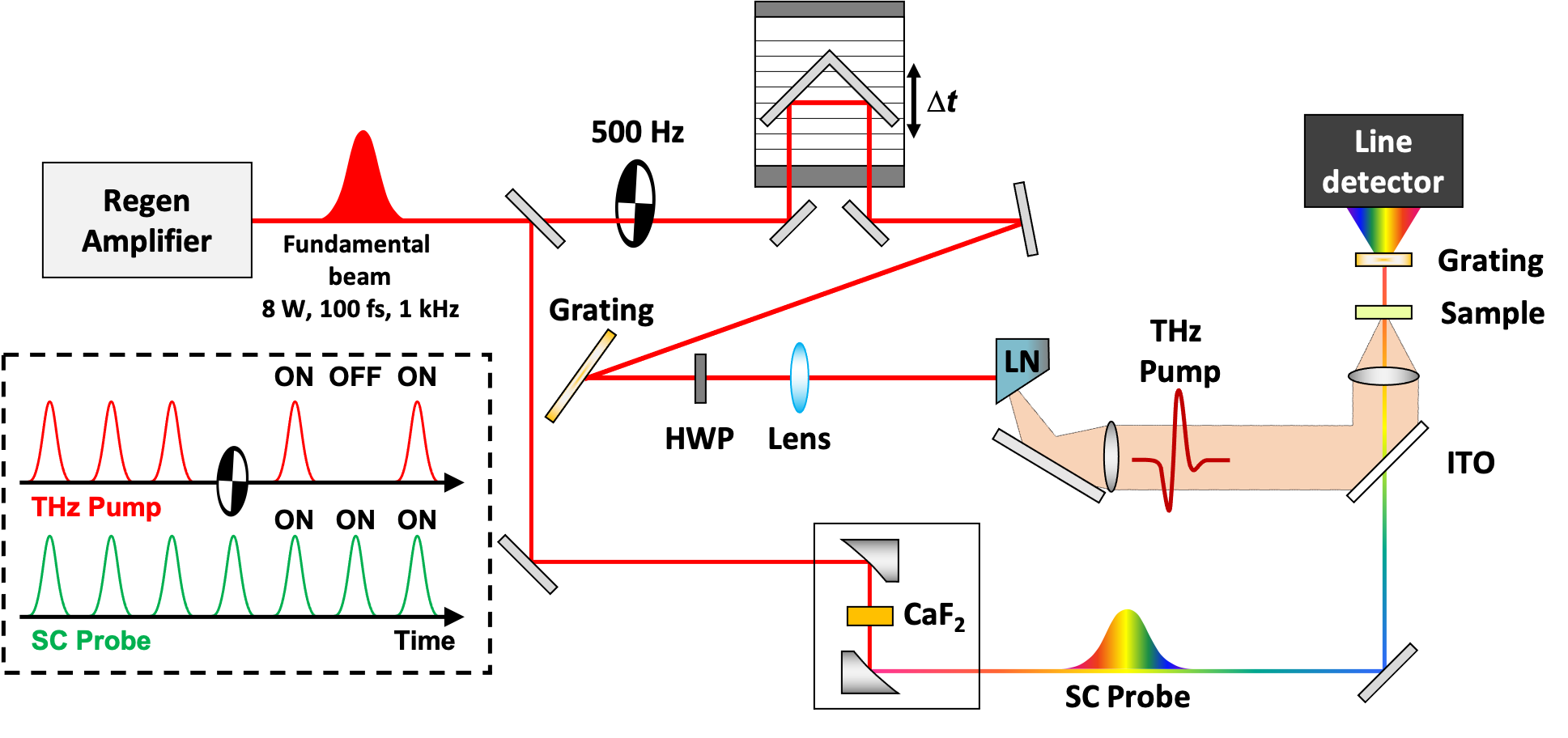}
\caption{\label{fig:SchematicExperiment} \textbf{THz Stark spectroscopy.} High-field single-cycle THz waveforms were generated via tilted-pulse-front pumping in LiNbO$_3$ and the fs-SC probe pulses were generated in CaF$_2$. The probe pulses passed collinearly with the THz pulses through the sample and were analyzed by a spectrometer. A chopper running at 500~Hz alternated between THz waveform on and off.}
\end{figure}

Figure~\ref{fig:SchematicExperiment} shows a schematic of the experimental setup. The high-field single-cycle THz waveforms were generated via tilted-pulse-front pumping in LiNbO$_3$ and the time-delayed femtosecond supercontinuum (fs-SC) probe pulses came from white-light generation in a 5-mm-thick CaF$_2$ crystal, which was mounted in a continuously moving mount in order to avoid photo-darkening. In detail, a 1~kHz Ti:sapphire regenerative amplifier (Legend Elite Duo Femto, Coherent) delivering 90~fs pulses with an average power of 8~W at 800~nm was used to produce the single-cycle THz waveforms by optical rectification in a prism-cut LiNbO$_3$ crystal \cite{Hebling2002, Fulop2014}. The THz waveforms were imaged to the sample position by a combination of two lenses with focal lengths of 100~mm and 50~mm, resulting in a 2:1 demagnification. 

THz waveform and fs-SC probe pulses were combined collinearly with an indium tin oxide (ITO) coated glass slide, which acts as a mirror at THz frequencies but is transparent at optical frequencies. The relative polarization between the THz waveform and the probe pulse was adjusted by an achromatic half-wave plate in the probe arm and both were subsequently focused to the sample using a TPX lens. The probe beam waist was set to $w_\mathrm{SC}$~$\approx$~17~$\mu$m, which is more than one order of magnitude smaller than the THz beam waist of $w_\mathrm{THz}$~$\approx$~1~mm, and thus probes an area of almost constant THz electric field strength at the center of the THz spot. After the sample, the probe spectrum was analyzed with a 1024-pixel CMOS array (Glaz-I, Synertronic Designs) on a pulse-to-pulse basis. A phase-locked chopper blocked every other THz waveform and from two consecutive probe spectra (with and without THz) the change in absorption, i.e. $\Delta A(\lambda) = A_\text{THz on}(\lambda) - A_\text{THz off}(\lambda) = -\log_{10}(T_\text{THz on}(\lambda) / T_\text{THz off}(\lambda))$, was calculated as a function of wavelength $\lambda$. In order to realize a sufficiently high signal-to-noise ratio we typically averaged more than 5000 pulse pairs. From a number of such measurements for different time delays $\tau$ between the THz waveform and the probe pulse, we construct two-dimensional color-coded Stark maps $\Delta A(\tau,\lambda)$ as shown in Fig.~\ref{fig:GVD_correction}a.

\subsubsection{Group velocity dispersion correction}

\begin{figure} [ht!]
\centering
\includegraphics[width=0.8\columnwidth]{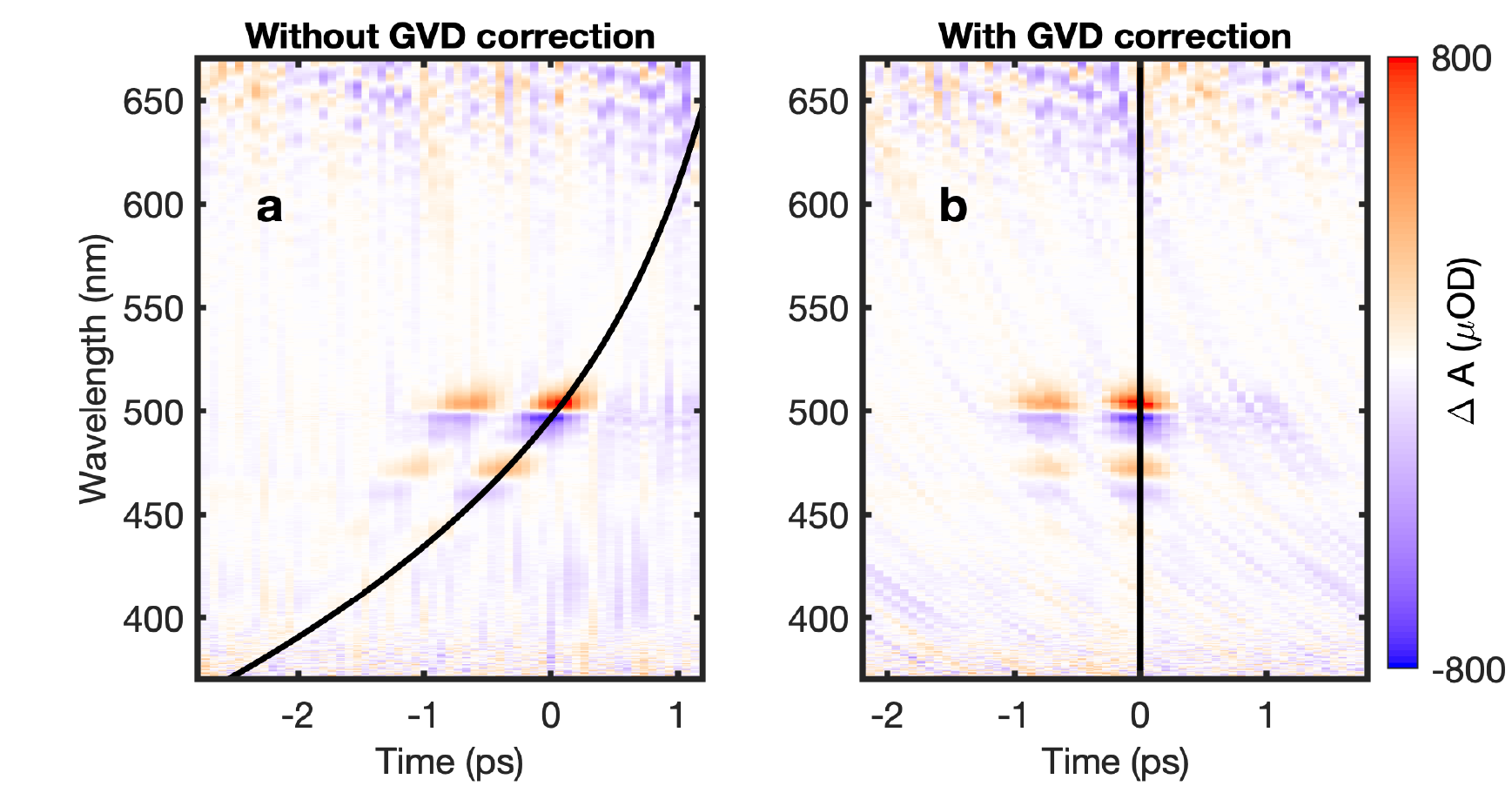}
\caption{\label{fig:GVD_correction} Measured color-coded Stark map for anthanthrene versus time delay and wavelength before \textbf{a} and after \textbf{b} GVD correction. The black solid curve indicates the fitted 3rd order polynomial representing the GVD, which turns into a vertical line after GVD correction.}
\end{figure}

The fs-SC probe pulses experience group velocity dispersion (GVD) due to a number of dispersive optical elements through which they have to propagate. As a result, each spectral component has a different effective time delay with respect to the THz waveform. Since the total GVD is independent of sample and polarization, it can be corrected for. As an example, Fig.~\ref{fig:GVD_correction}a shows the measured and color-coded Stark signals (anthanthrene) versus time delay between probe pulse and THz waveform and wavelength of probe pulse. The black solid curve shows the 3rd order polynomial fitting, which is subsequently used for GVD correction. The GVD correction was confirmed by measuring the cross-phase modulation in a quartz substrate at the sample position. In essence, all rows of data are shifted by a time delay that is given by the 3rd order polynomial curve and Fig.~\ref{fig:GVD_correction}b shows the data after GVD correction. All measurements on the solid curve shown in Fig.~\ref{fig:GVD_correction}a now have the same corrected time delay.

\subsubsection{Background measurements}

\begin{figure} [ht!]
\centering
\includegraphics[width=0.8\columnwidth]{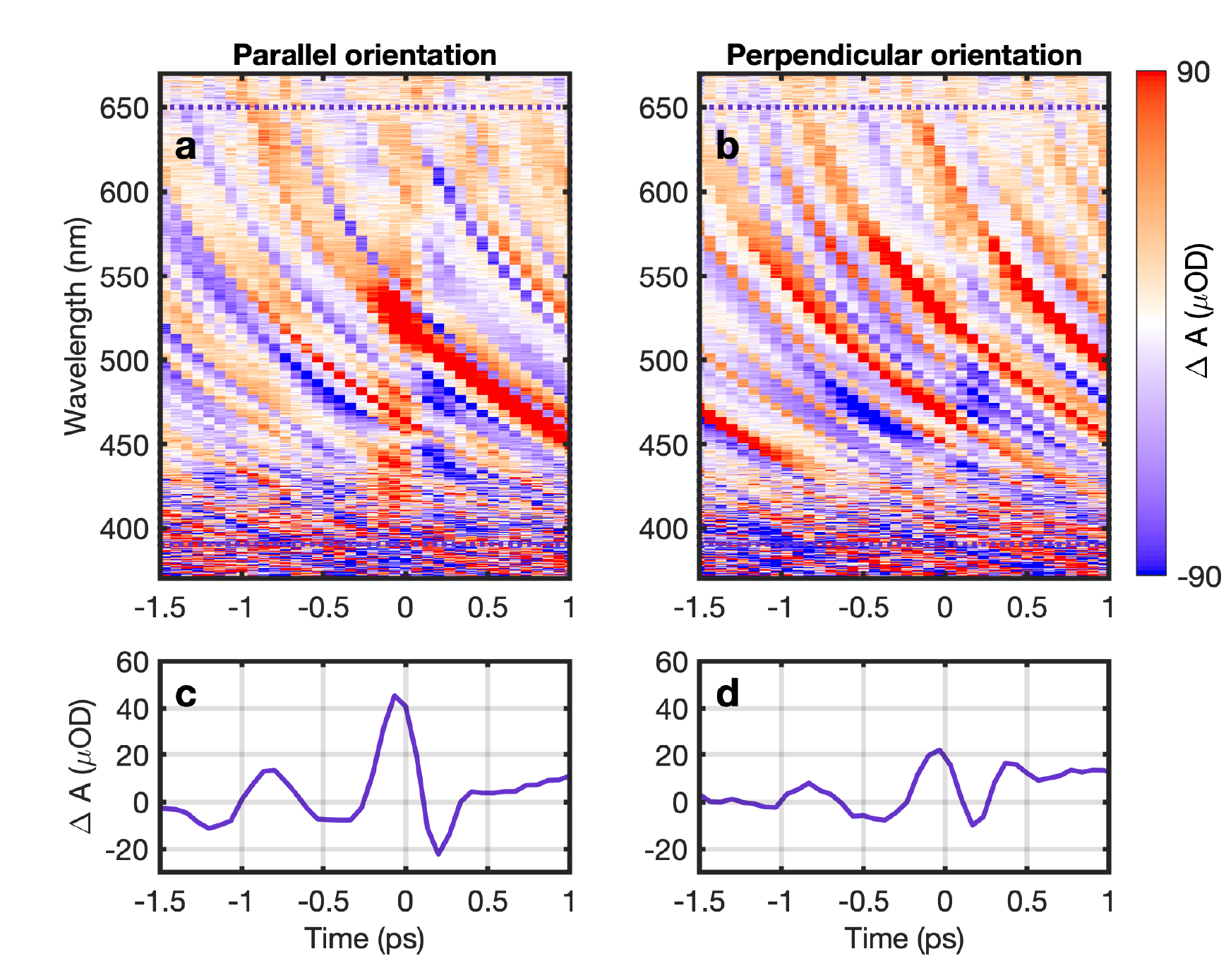}
\caption{\label{fig:Toluene_bothpol} Background measurement with pure toluene for \textbf{a} parallel and \textbf{b} perpendicular orientation of THz and probe pulses. Spectral average of the change in absorption between the purple dotted lines as a function of time delay for \textbf{c} parallel and \textbf{d} perpendicular orientation.}
\end{figure}

To characterize the measurement background, we recorded signals for the pure solvent. Figure~\ref{fig:Toluene_bothpol} shows the spectra for parallel and perpendicular orientation of THz and probe polarization. For all time delays, the measurement is dominated by random noise. Note that the GVD correction introduces an apparent parabolic structure. Around time delay zero we find a positive signal for all wavelength, which is attributed to a THz Kerr effect observed in low-polar liquids \cite{Sajadi2017}. In Fig.~\ref{fig:Toluene_bothpol}c,d the two signals integrated along the wavelength axis are shown as function of time delay. In agreement with theory, the signal for parallel orientation is larger than that for perpendicular orientation. For all measurements presented in this paper, we subtracted this background signal before plotting and analysing the data.

\subsubsection{Additional THz Stark spectroscopy results}

\begin{figure} [ht!]
\centering
\includegraphics[width=0.8\columnwidth]{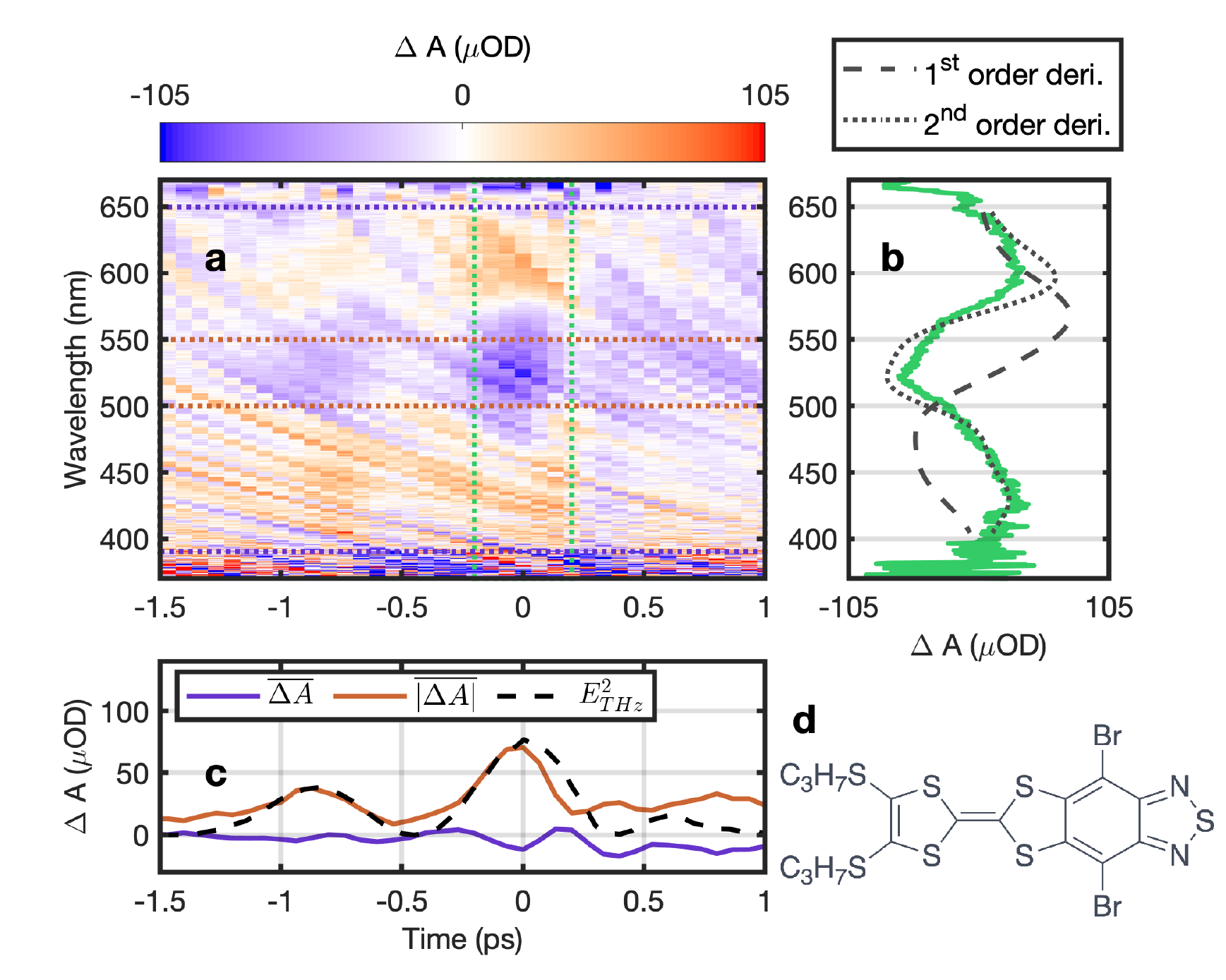}
\caption{\label{fig:TTF-BTD_perpol} \textbf{THz Stark signal of TTF-BTD with perpendicular orientation.} \textbf{a} False-color plot of the measured change in absorption spectrum as a function of time delay between THz and probe pulse and wavelength. \textbf{b} Time-averaged (between the two green dotted lines) change in absorption versus wavelength (green solid curve) compared to the scaled first (black dashed curve) and second order derivative (black dotted curve) of the ground state absorption spectrum. \textbf{c} Spectral average of the change in absorption between the purple dotted (purple curve) and red dotted lines (red curve) in \textbf{a}. The red curve is compared to the scaled square of the measured THz electric field $E_\mathrm{THz}^2$ (black dashed curve). \textbf{d} Chemical structure of TTF-BTD.} 
\end{figure}

\begin{figure} [ht!]
\centering
\includegraphics[width=0.8\columnwidth]{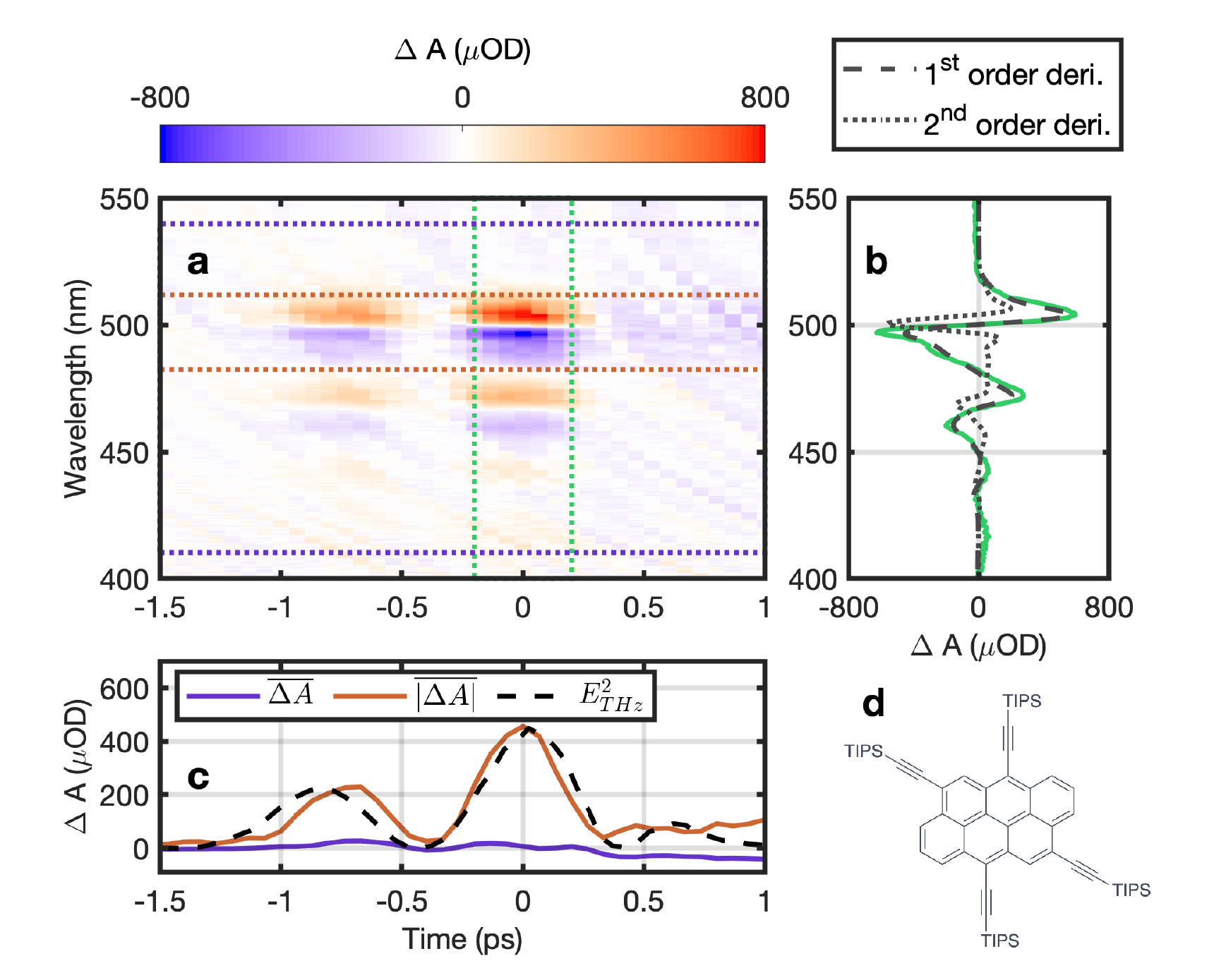}
\caption{\label{fig:Anthanthrene_parpol} \textbf{THz Stark signal of anthanthrene with perpendicular orientation.} \textbf{a} False-color plot of the measured change in absorption spectrum as a function of time delay between THz and probe pulse and wavelength. \textbf{b} Time-averaged (between the two green dotted lines) change in absorption versus wavelength (green solid curve) compared to the scaled first (black dashed curve) and second order derivative (black dotted curve) of the ground state absorption spectrum. \textbf{c} Spectral average of the change in absorption between the purple dotted (purple curve) and red dotted lines (red curve) in \textbf{a}. The red curve is compared to the scaled square of the measured THz electric field $E_\mathrm{THz}^2$ (black dashed curve). \textbf{d} Chemical structure of anthanthrene.}
\end{figure}

In the main text we only show results for parallel polarization between THz waveform and probe pulse. However, the Liptay analysis requires Stark signals for two different relative polarization orientations, ideally but not necessarily parallel and perpendicular. Hence, for completeness we here show the signals for perpendicular orientation. While Fig.~\ref{fig:TTF-BTD_perpol} shows the perpendicular orientation for TTF-BTD, Fig.~\ref{fig:Anthanthrene_parpol} shows the corresponding results for anthanthrene. 

\subsubsection{Characterization of THz pulses}

\begin{figure} [ht!]
\centering
\includegraphics[width=0.8\columnwidth]{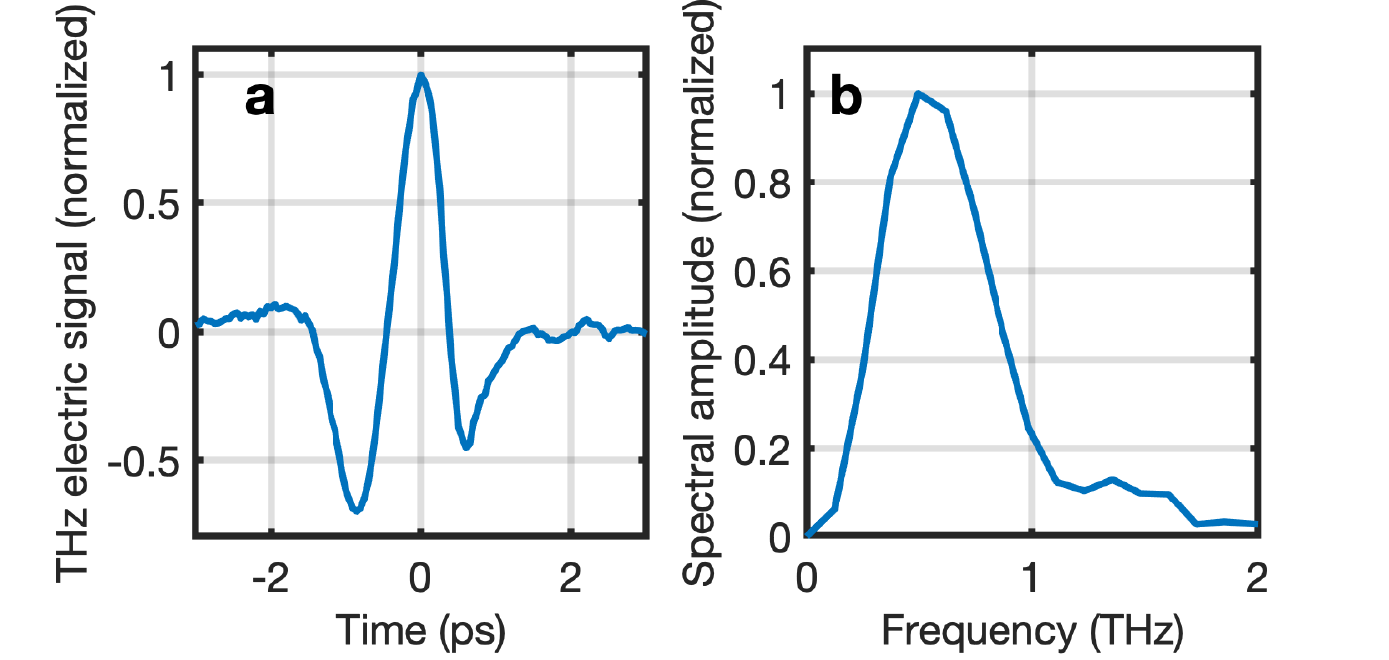}
\caption{\label{fig:THzPulse_and_spectrum} \textbf{a} Measured electro-optic signal at the sample position and \textbf{b} corresponding spectrum.}
\end{figure}

To determine the THz electric field strength in air we assume that the spatio-temporal electric field can be expressed in a product $E(x,y,t) = E_0 g_x(x) g_y(y) f(t)$, where $E_0$ is the peak electric field strength and $g_x(x)$, $g_y(y)$, and $f(t)$ are spatially and temporally dependent functions normalized to a peak value of one. The peak electric field strength $E_0$ was determined from three measurements, i.e. the average power $P_{avg}$ at a repetition rate of $f_{rep}$, the spatial profiles $g_x^2(x)$ and $g_y^2(y)$, and the time dependence $f(t)$ and is calculated via

\begin{equation}
\label{eq:THz_fieldstreng}
E_0 = \sqrt{\frac{P_{avg}}{\epsilon_0 c f_{rep} \int_{-\infty}^{\infty} g^2_x(x) dx \int_{-\infty}^{\infty} g^2_y(y) dy\int_{-\infty}^{\infty} f^2(t) dt}},
\end{equation}

where $\epsilon_0$ is the vacuum permittivity and $c$ is the speed of light in vacuum. The average power was recorded by a calibrated THz power meter (with a resolution of \SI{50}{\uW} and a relative error of $\pm 12$\%), the spatial profiles were extracted from two perpendicular knife-edge measurements, and the time dependence was measured by electro-optic sampling in a 0.3-mm-thick GaP $<110>$ crystal. Note that all measurements are taken at the sample position. The maximum THz power was measured to 3.6~mW at 1~kHz repetition rate, the THz spot was nearly Gaussian with a beam waist of 1~mm in x- as well as in y-direction, and the measured electro-optic signal \cite{Wu1995, Nahata1996, Brunner2014} and the corresponding spectrum are shown in Fig.~\ref{fig:THzPulse_and_spectrum}a and b. Inserting the three measurements in Eq.(\ref{eq:THz_fieldstreng}) results in a peak electric field of $E_0 = \SI[separate-uncertainty=true]{400 \pm 24}{\kV/\cm}$ in air.

The effective field strength, at which the probe pulse interrogates the molecular system, is typically smaller because of a number of effects. The dominating reduction factor comes from the geometry and material of the cuvette. For instance, Fresnel reflections or Fabry-Perot interferences at or within the cuvette limit the maximum electric field strength. Some solvents also have a no-negligible THz absorption coefficient and consequently reduce the electric field strength exponentially along the probe's propagation direction. Moreover, several effects lead to a smearing and averaging of the signal, such as the finite size of the probe interrogating the sample at different THz electric field strengths or the finite duration of the probe pulse and the group velocity mismatch between THz waveform and probe pulse. In order to account for all these effects we performed finite difference time-domain simulations propagating the THz waveform together with a time delayed probe pulse through the sample cuvette. We found that the time dependence of the THz electric field experienced by the probe pulse is almost identical to the free space THz waveform, however the peak electric field strength is reduced by a factor of 0.7 resulting in a maximum electric field of $E_0 = \SI[separate-uncertainty=true]{280 \pm 17}{\kV/\cm}$ in the sample.

\subsubsection{Liptay analysis}

The following analysis closely follows the recipe outlined in reference \cite{Liptay1969}. After having identified the Stark-active transitions, the Stark spectra are subsequently analyzed with the Liptay formalism (for details also see references \cite{Rohwer2018} and \cite{Bublitz1997a}). The analytic expression derived by Liptay links the molar absorption $\Delta \epsilon(\bar{\nu})$ as a function of wavenumber to the square of the electric field $\vb{E}$ and ground state absorption spectrum $\epsilon(\bar{\nu})$

\begin{equation}
   \Delta \epsilon(\bar{\nu}) = f_l^2 \abs{\vb{E}}^2 \left\{ A_{\chi} \epsilon(\bar{\nu})  + \frac{B_{\chi}}{15 h c} \bar{\nu} \dv{\bar{\nu}} \left( \frac{\epsilon(\bar{\nu})}{\bar{\nu}} \right) + \frac{C_{\chi}}{30 h^2 c^2} \bar{\nu} \dv[2]{\bar{\nu}} \left( \frac{\epsilon(\bar{\nu})}{\bar{\nu}} \right)\right\},
\end{equation}

where $h$ is Planck's constant and $c$ the speed of light. The model assumes an isotropic distribution of transition dipole moments, which is provided by freezing the sample in conventional Stark spectroscopy. The measured Stark spectra are fitted with a weighted combination of the zeroth, first, and second order derivative of the ground state absorption spectrum. The coefficient $A_{\chi}$ is determined by the transition polarizability and/or the transition hyperpolarizability of the sample, which can usually be neglected for immobilized samples. The second and third coefficients are given by

\begin{align}
    B_{\chi} &= \frac{5}{2} \text{Tr}(\underline{\Delta \alpha}) + \left( 3 \cos^2 \chi - 1 \right)\left(\frac{3}{2}\vb{m} \underline{\Delta \alpha} \vb{m} - \frac{1}{2} \text{Tr}(\underline{\Delta \alpha}) \right) \\
    C_{\chi} &= \abs{\Delta \mu}^2 \left\{ 5 + \left( 3 \cos^2\chi - 1\right)\left(3 \cos^2 \zeta - 1\right) \right\},
\end{align}

where $\text{Tr}(\underline{\Delta \alpha})$ is the trace of the polarizability tensor, $\vb{m} \underline{\Delta \alpha} \vb{m}$ is its projection along the transition dipole moment, $\chi$ is the angle between the applied electric field and the probe polarization, $\Delta \mu$ is the change in dipole moment and $\zeta$ is the angle between the change in dipole moment and the transition dipole moment.

\begin{figure} [ht!]
\centering
\includegraphics[width=0.8\columnwidth]{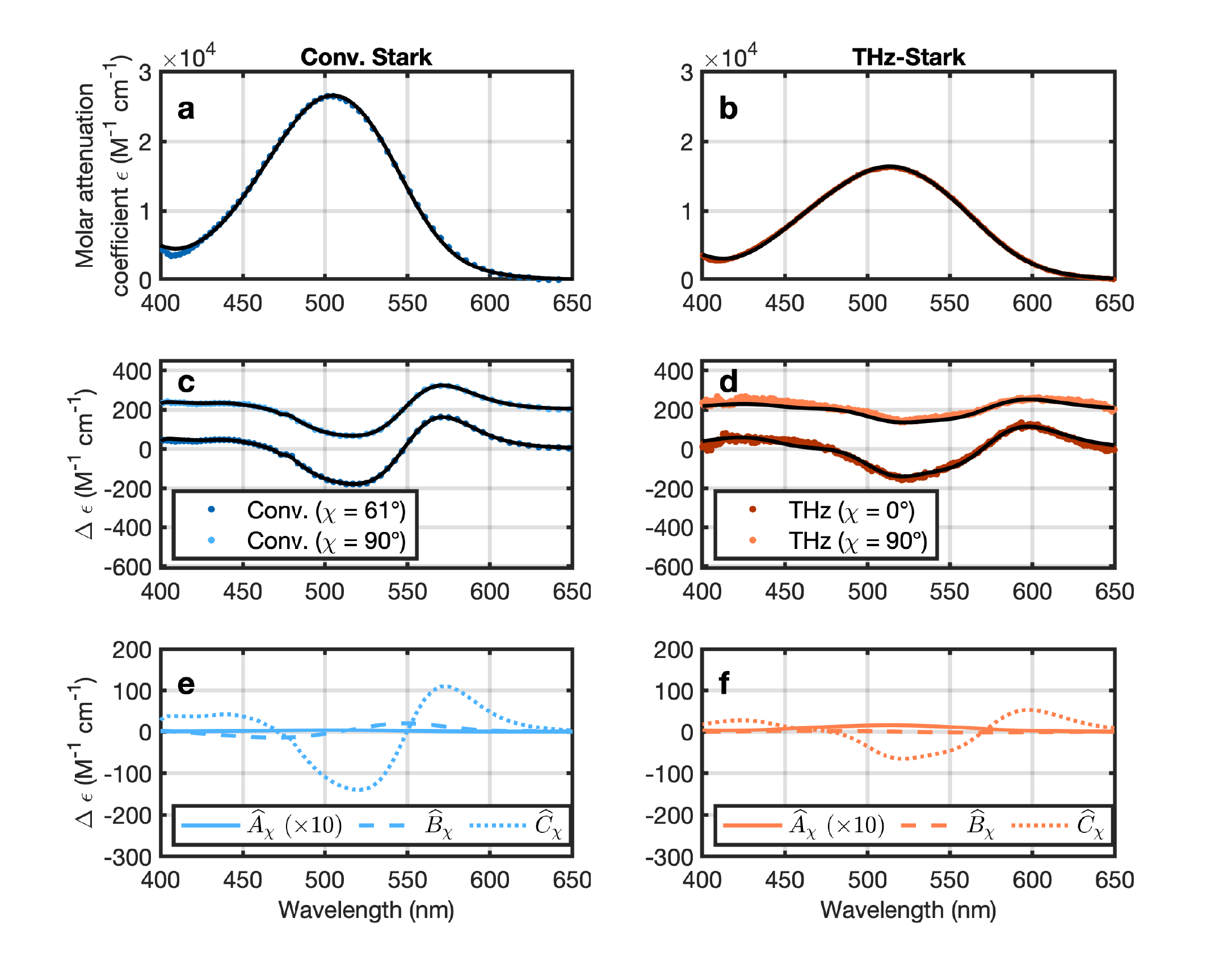}
\caption{\label{fig:StarkFits_TTF-BTD} \textbf{TTF-BTD conventional Stark and THz-Stark spectra fitting} \textbf{a, b} Ground-state absorption spectrum $\epsilon$ of TTF-BTD sample at 77~K (\textbf{a}) and at 295~K (\textbf{b}). The dots represent the data points and the black solid curves represents the fits. \textbf{c, d} Measured Stark spectra for two different incidence angles (dots) and corresponding fits (black solid curves) for the conventional Stark measurement \textbf{c} and THz-Stark measurement  \textbf{d}. For better visualization, the curves for $\chi = 90^\circ$ are arbitrarily shifted along the $\Delta \epsilon$-axis. \textbf{e, f} Contribution of the zeroth (solid curve), first (dashed curve) and second (dotted curve) order derivative line form for the Stark spectra for $\chi = 90^\circ$.}
\end{figure}

The molecular parameters are extracted by simultaneously fitting the ground-state absorption spectra and two Stark spectra for different probe polarizations. Figure~\ref{fig:StarkFits_TTF-BTD} shows the measured data (dotted curves) and the corresponding fits (black solid curves) for the TTF-BTD sample. Figure~\ref{fig:StarkFits_TTF-BTD}a and Fig.~\ref{fig:StarkFits_TTF-BTD}c show the results for the conventional Stark measurement at 77~K, while Fig.~\ref{fig:StarkFits_TTF-BTD}b and Fig.~\ref{fig:StarkFits_TTF-BTD}d show the results for the THz-Stark measurement at 295~K (same data as shown in Fig.~4 in the main text). For both experiments we find reasonable agreement between the fits and the measured data with a slightly better fit quality for the conventional Stark measurements. Figure~\ref{fig:StarkFits_TTF-BTD}e and Fig.~\ref{fig:StarkFits_TTF-BTD}f separately show the contribution of the zeroth, first and second order contribution to the fitted Stark signal for $\chi = 90^\circ$. The contribution of the zeroth order $\widehat{A}_{\chi}$ is multiplied by 10 and we a find negligible contribution to the Stark signal for both the conventional and the THz experiment, which confirms that the alignment of molecules due to the electric field is negligible. 

~~~

\begin{figure} [ht!]
\centering
\includegraphics[width=0.8\columnwidth]{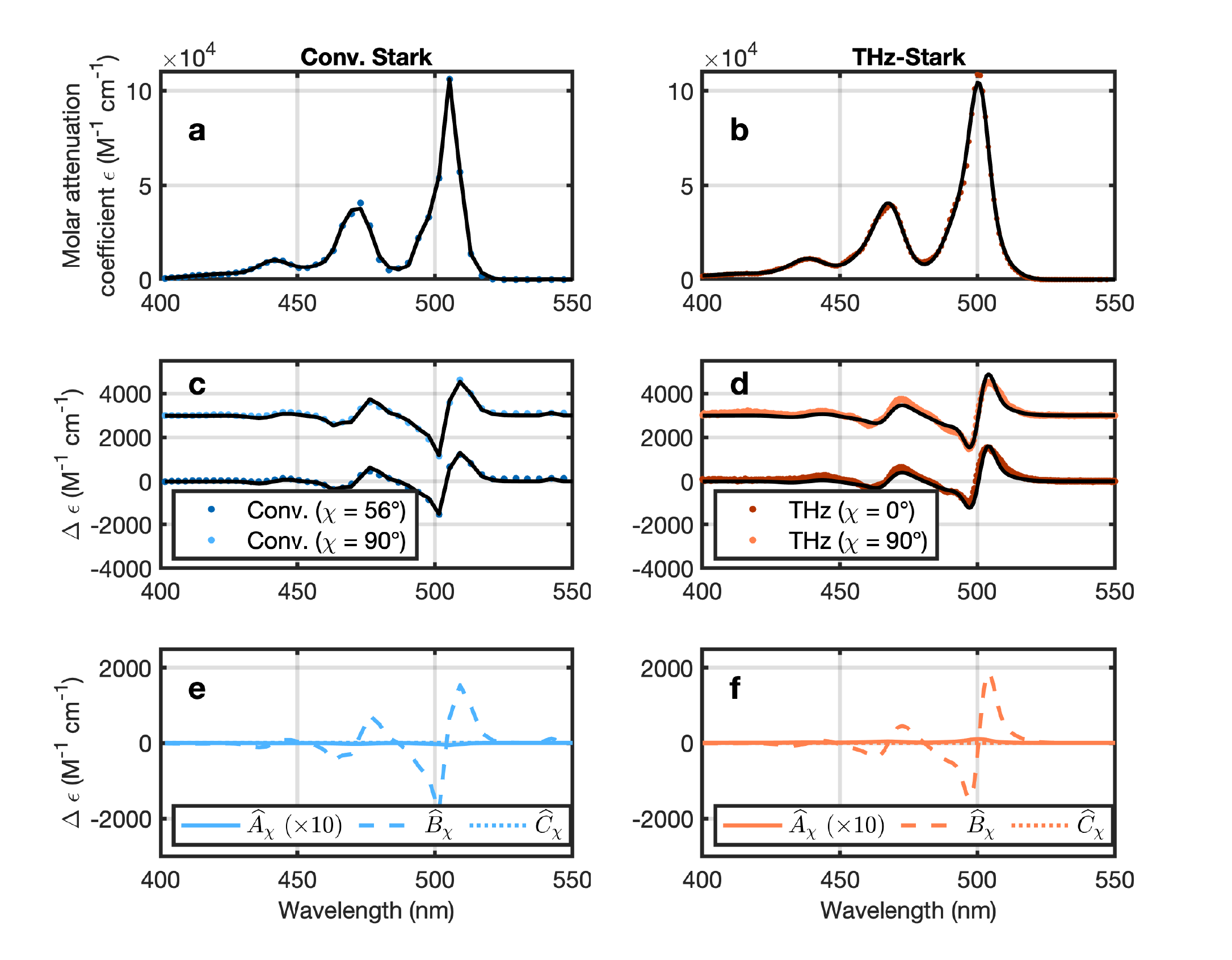}
\caption{\label{fig:StarkFits_Anthanthrene}  \textbf{Anthanthrene conventional Stark and THz-Stark spectra fitting} \textbf{a, b} Ground-state absorption spectrum $\epsilon$ of anthanthrene sample at 77~K (\textbf{a}) and at 295~K (\textbf{b}). The dots represent the data points and the black solid curves represents the fits. \textbf{c, d} Measured Stark spectra for two different incidence angles (dots) and corresponding fits (black solid curves) for the conventional Stark measurement \textbf{c} and THz-Stark measurement  \textbf{d}. For better visualization, the curves for $\chi = 90^\circ$ are arbitrarily shifted along the $\Delta \epsilon$-axis. \textbf{e, f} Contribution of the zeroth (solid curve), first (dashed curve) and second (dotted curve) order derivative line form for the Stark spectra for $\chi = 90^\circ$.}
\end{figure}

The same fitting results are shown in Fig.~\ref{fig:StarkFits_Anthanthrene} for the anthanthrene sample. Also here we find reasonable agreement between the fits and the measured data for both conventional and THz Stark spectroscopy.

\subsubsection{Local field correction factor} 

The local field correction factor gives a measure of how the solvent cavity affects the field inside the cavity when an external electric field is applied. The calculations were done in analogy to those described in literature \cite{Aubret2019, Stanley2001, Premvardhan1998}. We approximated the molecule as occupying a cavity with an ellipsoidal shape. For anthanthrene we estimate the ellipsoid axes to 15~\AA, 15~\AA, and 3~\AA, while for the TTB-BTD we estimate them to 15~\AA, 7~\AA, and 3~\AA. Note that reasonable variation of these parameters has only minor effects to the local field correction factor. Based on literature, the dielectric constant of toluene at 77~K and zero frequency is 2.52 \cite{Isnardi1922}, while at room temperature and 400~GHz it is 2.27 \cite{Ro/nne2000}. The local field correction factors $f_l$ for TTF-BTD and anthanthrene are estimated to be 1.30 and 1.33 respectively for conventional Stark spectroscopy and to be 1.26 and 1.29 respectively for THz Stark spectroscopy. Hence, the fitted values Tr($\Delta \alpha$) have to be divided by ${f_l}^2$ and $\Delta \mu$ by $f_l$ before the numbers can be compared with the DFT calculation.

\clearpage

\bibliography{library}


\begin{thebibliography}{71}
\ifx \bisbn   \undefined \def \bisbn  #1{ISBN #1}\fi
\ifx \binits  \undefined \def \binits#1{#1}\fi
\ifx \bauthor  \undefined \def \bauthor#1{#1}\fi
\ifx \batitle  \undefined \def \batitle#1{#1}\fi
\ifx \bjtitle  \undefined \def \bjtitle#1{#1}\fi
\ifx \bvolume  \undefined \def \bvolume#1{\textbf{#1}}\fi
\ifx \byear  \undefined \def \byear#1{#1}\fi
\ifx \bissue  \undefined \def \bissue#1{#1}\fi
\ifx \bfpage  \undefined \def \bfpage#1{#1}\fi
\ifx \blpage  \undefined \def \blpage #1{#1}\fi
\ifx \burl  \undefined \def \burl#1{\textsf{#1}}\fi
\ifx \doiurl  \undefined \def \doiurl#1{\url{https://doi.org/#1}}\fi
\ifx \betal  \undefined \def \betal{\textit{et al.}}\fi
\ifx \binstitute  \undefined \def \binstitute#1{#1}\fi
\ifx \binstitutionaled  \undefined \def \binstitutionaled#1{#1}\fi
\ifx \bctitle  \undefined \def \bctitle#1{#1}\fi
\ifx \beditor  \undefined \def \beditor#1{#1}\fi
\ifx \bpublisher  \undefined \def \bpublisher#1{#1}\fi
\ifx \bbtitle  \undefined \def \bbtitle#1{#1}\fi
\ifx \bedition  \undefined \def \bedition#1{#1}\fi
\ifx \bseriesno  \undefined \def \bseriesno#1{#1}\fi
\ifx \blocation  \undefined \def \blocation#1{#1}\fi
\ifx \bsertitle  \undefined \def \bsertitle#1{#1}\fi
\ifx \bsnm \undefined \def \bsnm#1{#1}\fi
\ifx \bsuffix \undefined \def \bsuffix#1{#1}\fi
\ifx \bparticle \undefined \def \bparticle#1{#1}\fi
\ifx \barticle \undefined \def \barticle#1{#1}\fi
\bibcommenthead
\ifx \bconfdate \undefined \def \bconfdate #1{#1}\fi
\ifx \botherref \undefined \def \botherref #1{#1}\fi
\ifx \url \undefined \def \url#1{\textsf{#1}}\fi
\ifx \bchapter \undefined \def \bchapter#1{#1}\fi
\ifx \bbook \undefined \def \bbook#1{#1}\fi
\ifx \bcomment \undefined \def \bcomment#1{#1}\fi
\ifx \oauthor \undefined \def \oauthor#1{#1}\fi
\ifx \citeauthoryear \undefined \def \citeauthoryear#1{#1}\fi
\ifx \endbibitem  \undefined \def \endbibitem {}\fi
\ifx \bconflocation  \undefined \def \bconflocation#1{#1}\fi
\ifx \arxivurl  \undefined \def \arxivurl#1{\textsf{#1}}\fi
\csname PreBibitemsHook\endcsname

\bibitem{Brunschwig1998}
\begin{barticle}
\bauthor{\bsnm{Brunschwig}, \binits{B.S.}},
\bauthor{\bsnm{Creutz}, \binits{C.}},
\bauthor{\bsnm{Sutin}, \binits{N.}}:
\batitle{{Electroabsorption spectroscopy of charge transfer states of
  transition metal complexes}}.
\bjtitle{Coord. Chem. Rev.}
\bvolume{177}(\bissue{1}),
\bfpage{61}--\blpage{79}
(\byear{1998}).
\doiurl{10.1016/S0010-8545(98)00188-X}
\end{barticle}
\endbibitem

\bibitem{Bublitz1997}
\begin{barticle}
\bauthor{\bsnm{Bublitz}, \binits{G.U.}},
\bauthor{\bsnm{Ortiz}, \binits{R.}},
\bauthor{\bsnm{Marder}, \binits{S.R.}},
\bauthor{\bsnm{Boxer}, \binits{S.G.}}:
\batitle{{Stark Spectroscopy of Donor/Acceptor Substituted Polyenes}}.
\bjtitle{J. Am. Chem. Soc.}
\bvolume{119}(\bissue{14}),
\bfpage{3365}--\blpage{3376}
(\byear{1997}).
\doiurl{10.1021/ja9640814}
\end{barticle}
\endbibitem

\bibitem{Iimori2016}
\begin{barticle}
\bauthor{\bsnm{Iimori}, \binits{T.}},
\bauthor{\bsnm{Ito}, \binits{R.}},
\bauthor{\bsnm{Ohta}, \binits{N.}},
\bauthor{\bsnm{Nakano}, \binits{H.}}:
\batitle{{Stark Spectroscopy of Rubrene. I. Electroabsorption Spectroscopy and
  Molecular Parameters}}.
\bjtitle{J. Phys. Chem. A}
\bvolume{120}(\bissue{25}),
\bfpage{4307}--\blpage{4313}
(\byear{2016}).
\doiurl{10.1021/acs.jpca.6b02625}
\end{barticle}
\endbibitem

\bibitem{Karki1998}
\begin{barticle}
\bauthor{\bsnm{Karki}, \binits{L.}},
\bauthor{\bsnm{Vance}, \binits{F.W.}},
\bauthor{\bsnm{Hupp}, \binits{J.T.}},
\bauthor{\bsnm{LeCours}, \binits{S.M.}},
\bauthor{\bsnm{Therien}, \binits{M.J.}}:
\batitle{{Electronic Stark Effect Studies of a Porphyrin-Based Push-Pull
  Chromophore Displaying a Large First Hyperpolarizability: State-Specific
  Contributions to $\beta$}}.
\bjtitle{J. Am. Chem. Soc.}
\bvolume{120}(\bissue{11}),
\bfpage{2606}--\blpage{2611}
(\byear{1998}).
\doiurl{10.1021/ja973593v}
\end{barticle}
\endbibitem

\bibitem{Liptay1969}
\begin{barticle}
\bauthor{\bsnm{Liptay}, \binits{W.}}:
\batitle{{Electrochromism and Solvatochromism}}.
\bjtitle{Angew. Chemie Int. Ed. English}
\bvolume{8}(\bissue{3}),
\bfpage{177}--\blpage{188}
(\byear{1969}).
\doiurl{10.1002/anie.196901771}
\end{barticle}
\endbibitem

\bibitem{Locknar1998}
\begin{barticle}
\bauthor{\bsnm{Locknar}, \binits{S.A.}},
\bauthor{\bsnm{Peteanu}, \binits{L.A.}}:
\batitle{{Investigation of the Relationship between Dipolar Properties and
  Cis-Trans Configuration in Retinal Polyenes: A Comparative Study Using Stark
  Spectroscopy and Semiempirical Calculations}}.
\bjtitle{J. Phys. Chem. B}
\bvolume{102}(\bissue{21}),
\bfpage{4240}--\blpage{4246}
(\byear{1998}).
\doiurl{10.1021/jp980562t}
\end{barticle}
\endbibitem

\bibitem{Mathies1976}
\begin{barticle}
\bauthor{\bsnm{Mathies}, \binits{R.}},
\bauthor{\bsnm{Stryer}, \binits{L.}}:
\batitle{{Retinal has a highly dipolar vertically excited singlet state:
  implications for vision}}.
\bjtitle{Proc. Natl. Acad. Sci. U. S. A.}
\bvolume{73}(\bissue{7}),
\bfpage{2169}--\blpage{2173}
(\byear{1976}).
\doiurl{10.1073/pnas.73.7.2169}
\end{barticle}
\endbibitem

\bibitem{Pauszek2013}
\begin{barticle}
\bauthor{\bsnm{Pauszek}, \binits{R.F.}},
\bauthor{\bsnm{Kodali}, \binits{G.}},
\bauthor{\bsnm{Caldwell}, \binits{S.T.}},
\bauthor{\bsnm{Fitzpatrick}, \binits{B.}},
\bauthor{\bsnm{Zainalabdeen}, \binits{N.Y.}},
\bauthor{\bsnm{Cooke}, \binits{G.}},
\bauthor{\bsnm{Rotello}, \binits{V.M.}},
\bauthor{\bsnm{Stanley}, \binits{R.J.}}:
\batitle{{Excited State Charge Redistribution and Dynamics in the
  Donor-$\pi$-Acceptor Flavin Derivative ABFL}}.
\bjtitle{J. Phys. Chem. B}
\bvolume{117}(\bissue{49}),
\bfpage{15684}--\blpage{15694}
(\byear{2013}).
\doiurl{10.1021/jp406420h}
\end{barticle}
\endbibitem

\bibitem{Pein2019terahertz}
\begin{barticle}
\bauthor{\bsnm{Pein}, \binits{B.C.}},
\bauthor{\bsnm{Lee}, \binits{C.K.}},
\bauthor{\bsnm{Shi}, \binits{L.}},
\bauthor{\bsnm{Shi}, \binits{J.}},
\bauthor{\bsnm{Chang}, \binits{W.}},
\bauthor{\bsnm{Hwang}, \binits{H.Y.}},
\bauthor{\bsnm{Scherer}, \binits{J.}},
\bauthor{\bsnm{Coropceanu}, \binits{I.}},
\bauthor{\bsnm{Zhao}, \binits{X.}},
\bauthor{\bsnm{Zhang}, \binits{X.}},
\bauthor{\bsnm{Bulovi{\'{c}}}, \binits{V.}},
\bauthor{\bsnm{Bawendi}, \binits{M.G.}},
\bauthor{\bsnm{Willard}, \binits{A.P.}},
\bauthor{\bsnm{Nelson}, \binits{K.A.}}:
\batitle{{Terahertz-Driven Stark Spectroscopy of CdSe and CdSe–CdS
  Core–Shell Quantum Dots}}.
\bjtitle{Nano Lett.}
\bvolume{19}(\bissue{11}),
\bfpage{8125}--\blpage{8131}
(\byear{2019}).
\doiurl{10.1021/acs.nanolett.9b03342}
\end{barticle}
\endbibitem

\bibitem{Hopkins2003}
\begin{barticle}
\bauthor{\bsnm{Hopkins}, \binits{N.}},
\bauthor{\bsnm{Stanley}, \binits{R.J.}}:
\batitle{{Measurement of the Electronic Properties of the Flavoprotein Old
  Yellow Enzyme (OYE) and the OYE:p-Cl Phenol Charge-Transfer Complex Using
  Stark Spectroscopy}}.
\bjtitle{Biochemistry}
\bvolume{42}(\bissue{4}),
\bfpage{991}--\blpage{999}
(\byear{2003}).
\doiurl{10.1021/bi0268908}
\end{barticle}
\endbibitem

\bibitem{Oh1989}
\begin{barticle}
\bauthor{\bsnm{Oh}, \binits{D.H.}},
\bauthor{\bsnm{Boxer}, \binits{S.G.}}:
\batitle{{Stark effect spectra of Ru(diimine)32+ complexes}}.
\bjtitle{J. Am. Chem. Soc.}
\bvolume{111}(\bissue{3}),
\bfpage{1130}--\blpage{1131}
(\byear{1989}).
\doiurl{10.1021/ja00185a054}
\end{barticle}
\endbibitem

\bibitem{Reimers1991}
\begin{barticle}
\bauthor{\bsnm{Reimers}, \binits{J.R.}},
\bauthor{\bsnm{Hush}, \binits{N.S.}}:
\batitle{{Electronic properties of transition-metal complexes determined from
  electroabsorption (Stark) spectroscopy. 2. Mononuclear complexes of ruthenium
  (II)}}.
\bjtitle{J. Phys. Chem.}
\bvolume{95}(\bissue{24}),
\bfpage{9773}--\blpage{9781}
(\byear{1991})
\end{barticle}
\endbibitem

\bibitem{Shin1995}
\begin{barticle}
\bauthor{\bsnm{Shin}, \binits{Y.-g.K.}},
\bauthor{\bsnm{Brunschwig}, \binits{B.S.}},
\bauthor{\bsnm{Creutz}, \binits{C.}},
\bauthor{\bsnm{Sutin}, \binits{N.}}:
\batitle{{Toward a Quantitative Understanding of Dipole-Moment Changes in
  Charge-Transfer Transitions: Electroabsorption Spectroscopy of
  Transition-Metal Complexes}}.
\bjtitle{J. Am. Chem. Soc.}
\bvolume{117}(\bissue{33}),
\bfpage{8668}--\blpage{8669}
(\byear{1995}).
\doiurl{10.1021/ja00138a024}
\end{barticle}
\endbibitem

\bibitem{Roiati2014stark}
\begin{barticle}
\bauthor{\bsnm{Roiati}, \binits{V.}},
\bauthor{\bsnm{Mosconi}, \binits{E.}},
\bauthor{\bsnm{Listorti}, \binits{A.}},
\bauthor{\bsnm{Colella}, \binits{S.}},
\bauthor{\bsnm{Gigli}, \binits{G.}},
\bauthor{\bsnm{{De Angelis}}, \binits{F.}}:
\batitle{{Stark Effect in Perovskite/TiO2 Solar Cells: Evidence of Local
  Interfacial Order}}.
\bjtitle{Nano Lett.}
\bvolume{14}(\bissue{4}),
\bfpage{2168}--\blpage{2174}
(\byear{2014}).
\doiurl{10.1021/nl500544c}
\end{barticle}
\endbibitem

\bibitem{Mehata2012}
\begin{barticle}
\bauthor{\bsnm{Mehata}, \binits{M.S.}},
\bauthor{\bsnm{Hsu}, \binits{C.-S.}},
\bauthor{\bsnm{Lee}, \binits{Y.-P.}},
\bauthor{\bsnm{Ohta}, \binits{N.}}:
\batitle{{Electroabsorption and Electrophotoluminescence of
  Poly(2,3-diphenyl-5-hexyl-p-phenylene vinylene)}}.
\bjtitle{J. Phys. Chem. C}
\bvolume{116}(\bissue{28}),
\bfpage{14789}--\blpage{14795}
(\byear{2012}).
\doiurl{10.1021/jp302666f}
\end{barticle}
\endbibitem

\bibitem{Chowdhury2001}
\begin{barticle}
\bauthor{\bsnm{Chowdhury}, \binits{A.}},
\bauthor{\bsnm{Peteanu}, \binits{L.A.}},
\bauthor{\bsnm{Webb}, \binits{M.A.}},
\bauthor{\bsnm{Loppnow}, \binits{G.R.}}:
\batitle{{Stark spectroscopic studies of blue copper proteins: Azurin}}.
\bjtitle{J. Phys. Chem. B}
\bvolume{105}(\bissue{2}),
\bfpage{527}--\blpage{534}
(\byear{2001}).
\doiurl{10.1021/jp0025227}
\end{barticle}
\endbibitem

\bibitem{Kodali2009}
\begin{barticle}
\bauthor{\bsnm{Kodali}, \binits{G.}},
\bauthor{\bsnm{Siddiqui}, \binits{S.U.}},
\bauthor{\bsnm{Stanley}, \binits{R.J.}}:
\batitle{{Charge Redistribution in Oxidized and Semiquinone E. coli DNA
  Photolyase upon Photoexcitation: Stark Spectroscopy Reveals a Rationale for
  the Position of Trp382}}.
\bjtitle{J. Am. Chem. Soc.}
\bvolume{131}(\bissue{13}),
\bfpage{4795}--\blpage{4807}
(\byear{2009}).
\doiurl{10.1021/ja809214r}
\end{barticle}
\endbibitem

\bibitem{Lockhart1988}
\begin{barticle}
\bauthor{\bsnm{Lockhart}, \binits{D.J.}},
\bauthor{\bsnm{Boxer}, \binits{S.G.}}:
\batitle{{Stark effect spectroscopy of Rhodobacter sphaeroides and
  Rhodopseudomonas viridis reaction centers}}.
\bjtitle{Proc. Natl. Acad. Sci.}
\bvolume{85}(\bissue{1}),
\bfpage{107}--\blpage{111}
(\byear{1988}).
\doiurl{10.1073/pnas.85.1.107}
\end{barticle}
\endbibitem

\bibitem{Losche1987}
\begin{barticle}
\bauthor{\bsnm{L{\"{o}}sche}, \binits{M.}},
\bauthor{\bsnm{Feher}, \binits{G.}},
\bauthor{\bsnm{Okamura}, \binits{M.Y.}}:
\batitle{{The Stark effect in reaction centers from Rhodobacter sphaeroides
  R-26 and Rhodopseudomonas viridis.}}
\bjtitle{Proc. Natl. Acad. Sci.}
\bvolume{84}(\bissue{21}),
\bfpage{7537}--\blpage{7541}
(\byear{1987}).
\doiurl{10.1073/pnas.84.21.7537}
\end{barticle}
\endbibitem

\bibitem{Premvardhan2003}
\begin{barticle}
\bauthor{\bsnm{Premvardhan}, \binits{L.L.}},
\bauthor{\bparticle{van~der} \bsnm{Horst}, \binits{M.A.}},
\bauthor{\bsnm{Hellingwerf}, \binits{K.J.}},
\bauthor{\bparticle{van} \bsnm{Grondelle}, \binits{R.}}:
\batitle{{Stark spectroscopy on photoactive yellow protein, E46Q, and a
  nonisomerizing derivative, probes photo-induced charge motion}}.
\bjtitle{Biophys. J.}
\bvolume{84}(\bissue{5}),
\bfpage{3226}--\blpage{3239}
(\byear{2003})
\end{barticle}
\endbibitem

\bibitem{Somsen1998}
\begin{barticle}
\bauthor{\bsnm{Somsen}, \binits{O.J.G.}},
\bauthor{\bsnm{Chernyak}, \binits{V.}},
\bauthor{\bsnm{Frese}, \binits{R.N.}},
\bauthor{\bparticle{van} \bsnm{Grondelle}, \binits{R.}},
\bauthor{\bsnm{Mukamel}, \binits{S.}}:
\batitle{{Excitonic Interactions and Stark Spectroscopy of Light Harvesting
  Systems}}.
\bjtitle{J. Phys. Chem. B}
\bvolume{102}(\bissue{44}),
\bfpage{8893}--\blpage{8908}
(\byear{1998}).
\doiurl{10.1021/jp981114o}
\end{barticle}
\endbibitem

\bibitem{Verma2020}
\begin{barticle}
\bauthor{\bsnm{Verma}, \binits{N.}},
\bauthor{\bsnm{Tao}, \binits{Y.}},
\bauthor{\bsnm{Zou}, \binits{W.}},
\bauthor{\bsnm{Chen}, \binits{X.}},
\bauthor{\bsnm{Chen}, \binits{X.}},
\bauthor{\bsnm{Freindorf}, \binits{M.}},
\bauthor{\bsnm{Kraka}, \binits{E.}}:
\batitle{{A Critical Evaluation of Vibrational Stark Effect (VSE) Probes with
  the Local Vibrational Mode Theory}}.
\bjtitle{Sensors}
\bvolume{20}(\bissue{8}),
\bfpage{2358}
(\byear{2020}).
\doiurl{10.3390/s20082358}
\end{barticle}
\endbibitem

\bibitem{Bublitz1997a}
\begin{barticle}
\bauthor{\bsnm{Bublitz}, \binits{G.U.}},
\bauthor{\bsnm{Boxer}, \binits{S.G.}}:
\batitle{{STARK SPECTROSCOPY: Applications in Chemistry, Biology, and Materials
  Science}}.
\bjtitle{Annu. Rev. Phys. Chem.}
\bvolume{48}(\bissue{1}),
\bfpage{213}--\blpage{242}
(\byear{1997}).
\doiurl{10.1146/annurev.physchem.48.1.213}
\end{barticle}
\endbibitem

\bibitem{Keiber2016electro}
\begin{barticle}
\bauthor{\bsnm{Keiber}, \binits{S.}},
\bauthor{\bsnm{Sederberg}, \binits{S.}},
\bauthor{\bsnm{Schwarz}, \binits{A.}},
\bauthor{\bsnm{Trubetskov}, \binits{M.}},
\bauthor{\bsnm{Pervak}, \binits{V.}},
\bauthor{\bsnm{Krausz}, \binits{F.}},
\bauthor{\bsnm{Karpowicz}, \binits{N.}}:
\batitle{{Electro-optic sampling of near-infrared waveforms}}.
\bjtitle{Nat. Photonics}
\bvolume{10}(\bissue{3}),
\bfpage{159}--\blpage{162}
(\byear{2016}).
\doiurl{10.1038/nphoton.2015.269}
\end{barticle}
\endbibitem

\bibitem{Knorr2017phase}
\begin{barticle}
\bauthor{\bsnm{Knorr}, \binits{M.}},
\bauthor{\bsnm{Raab}, \binits{J.}},
\bauthor{\bsnm{Tauer}, \binits{M.}},
\bauthor{\bsnm{Merkl}, \binits{P.}},
\bauthor{\bsnm{Peller}, \binits{D.}},
\bauthor{\bsnm{Wittmann}, \binits{E.}},
\bauthor{\bsnm{Riedle}, \binits{E.}},
\bauthor{\bsnm{Lange}, \binits{C.}},
\bauthor{\bsnm{Huber}, \binits{R.}}:
\batitle{{Phase-locked multi-terahertz electric fields exceeding 13 MV/cm at a
  190 kHz repetition rate}}.
\bjtitle{Opt. Lett.}
\bvolume{42}(\bissue{21}),
\bfpage{4367}--\blpage{4370}
(\byear{2017}).
\doiurl{10.1364/OL.42.004367}
\end{barticle}
\endbibitem

\bibitem{Jones1995}
\begin{barticle}
\bauthor{\bsnm{Jones}, \binits{H.M.}},
\bauthor{\bsnm{Kunhardt}, \binits{E.E.}}:
\batitle{{Pulsed dielectric breakdown of pressurized water and salt
  solutions}}.
\bjtitle{J. Appl. Phys.}
\bvolume{77}(\bissue{2}),
\bfpage{795}--\blpage{805}
(\byear{1995}).
\doiurl{10.1063/1.359002}
\end{barticle}
\endbibitem

\bibitem{Krasucki1966}
\begin{barticle}
\bauthor{\bsnm{Krasucki}, \binits{Z.}},
\bauthor{\bsnm{Bowden}, \binits{F.P.}}:
\batitle{{Breakdown of liquid dielectrics}}.
\bjtitle{Proc. R. Soc. London. Ser. A. Math. Phys. Sci.}
\bvolume{294}(\bissue{1438}),
\bfpage{393}--\blpage{404}
(\byear{1966}).
\doiurl{10.1098/rspa.1966.0214}
\end{barticle}
\endbibitem

\bibitem{Zhang2018segmented}
\begin{barticle}
\bauthor{\bsnm{Zhang}, \binits{D.}},
\bauthor{\bsnm{Fallahi}, \binits{A.}},
\bauthor{\bsnm{Hemmer}, \binits{M.}},
\bauthor{\bsnm{Wu}, \binits{X.}},
\bauthor{\bsnm{Fakhari}, \binits{M.}},
\bauthor{\bsnm{Hua}, \binits{Y.}},
\bauthor{\bsnm{Cankaya}, \binits{H.}},
\bauthor{\bsnm{Calendron}, \binits{A.-L.}},
\bauthor{\bsnm{Zapata}, \binits{L.E.}},
\bauthor{\bsnm{Matlis}, \binits{N.H.}},
\bauthor{\bsnm{K{\"{a}}rtner}, \binits{F.X.}}:
\batitle{{Segmented terahertz electron accelerator and manipulator (STEAM)}}.
\bjtitle{Nat. Photonics}
\bvolume{12}(\bissue{6}),
\bfpage{336}--\blpage{342}
(\byear{2018}).
\doiurl{10.1038/s41566-018-0138-z}
\end{barticle}
\endbibitem

\bibitem{Amacher2014}
\begin{barticle}
\bauthor{\bsnm{Amacher}, \binits{A.}},
\bauthor{\bsnm{Luo}, \binits{H.}},
\bauthor{\bsnm{Liu}, \binits{Z.}},
\bauthor{\bsnm{Bircher}, \binits{M.}},
\bauthor{\bsnm{Cascella}, \binits{M.}},
\bauthor{\bsnm{Hauser}, \binits{J.}},
\bauthor{\bsnm{Decurtins}, \binits{S.}},
\bauthor{\bsnm{Zhang}, \binits{D.}},
\bauthor{\bsnm{Liu}, \binits{S.-X.}}:
\batitle{{Electronic tuning effects via cyano substitution of a fused
  tetrathiafulvalene–benzothiadiazole dyad for ambipolar transport
  properties}}.
\bjtitle{RSC Adv.}
\bvolume{4}(\bissue{6}),
\bfpage{2873}--\blpage{2878}
(\byear{2014}).
\doiurl{10.1039/C3RA46784H}
\end{barticle}
\endbibitem

\bibitem{Pop2013a}
\begin{barticle}
\bauthor{\bsnm{Pop}, \binits{F.}},
\bauthor{\bsnm{Amacher}, \binits{A.}},
\bauthor{\bsnm{Avarvari}, \binits{N.}},
\bauthor{\bsnm{Ding}, \binits{J.}},
\bauthor{\bsnm{Daku}, \binits{L.M.L.}},
\bauthor{\bsnm{Hauser}, \binits{A.}},
\bauthor{\bsnm{Koch}, \binits{M.}},
\bauthor{\bsnm{Hauser}, \binits{J.}},
\bauthor{\bsnm{Liu}, \binits{S.-X.}},
\bauthor{\bsnm{Decurtins}, \binits{S.}}:
\batitle{{Tetrathiafulvalene-Benzothiadiazoles as Redox-Tunable
  Donor–Acceptor Systems: Synthesis and Photophysical Study}}.
\bjtitle{Chem. – A Eur. J.}
\bvolume{19}(\bissue{7}),
\bfpage{2504}--\blpage{2514}
(\byear{2013}).
\doiurl{10.1002/chem.201202742}
\end{barticle}
\endbibitem

\bibitem{Giguere2013}
\begin{barticle}
\bauthor{\bsnm{Gigu{\`{e}}re}, \binits{J.-B.}},
\bauthor{\bsnm{Verolet}, \binits{Q.}},
\bauthor{\bsnm{Morin}, \binits{J.-F.}}:
\batitle{{4,10-Dibromoanthanthrone as a New Building Block for p-Type, n-Type,
  and Ambipolar $\pi$-Conjugated Materials}}.
\bjtitle{Chem. – A Eur. J.}
\bvolume{19}(\bissue{1}),
\bfpage{372}--\blpage{381}
(\byear{2013}).
\doiurl{10.1002/chem.201202878}
\end{barticle}
\endbibitem

\bibitem{Rohwer2018}
\begin{barticle}
\bauthor{\bsnm{Rohwer}, \binits{E.J.}},
\bauthor{\bsnm{Akbarimoosavi}, \binits{M.}},
\bauthor{\bsnm{Meckel}, \binits{S.E.}},
\bauthor{\bsnm{Liu}, \binits{X.}},
\bauthor{\bsnm{Geng}, \binits{Y.}},
\bauthor{\bsnm{{Lawson Daku}}, \binits{L.M.}},
\bauthor{\bsnm{Hauser}, \binits{A.}},
\bauthor{\bsnm{Cannizzo}, \binits{A.}},
\bauthor{\bsnm{Decurtins}, \binits{S.}},
\bauthor{\bsnm{Stanley}, \binits{R.J.}},
\bauthor{\bsnm{Liu}, \binits{S.-X.}},
\bauthor{\bsnm{Feurer}, \binits{T.}}:
\batitle{{Dipole Moment and Polarizability of Tunable Intramolecular Charge
  Transfer States in Heterocyclic $\pi$-Conjugated Molecular Dyads Determined
  by Computational and Stark Spectroscopic Study}}.
\bjtitle{J. Phys. Chem. C}
\bvolume{122}(\bissue{17}),
\bfpage{9346}--\blpage{9355}
(\byear{2018}).
\doiurl{10.1021/acs.jpcc.8b02268}
\end{barticle}
\endbibitem

\bibitem{Bublitz1998}
\begin{barticle}
\bauthor{\bsnm{Bublitz}, \binits{G.U.}},
\bauthor{\bsnm{Boxer}, \binits{S.G.}}:
\batitle{{Effective Polarity of Frozen Solvent Glasses in the Vicinity of
  Dipolar Solutes}}.
\bjtitle{J. Am. Chem. Soc.}
\bvolume{120}(\bissue{16}),
\bfpage{3988}--\blpage{3992}
(\byear{1998}).
\doiurl{10.1021/ja971665c}
\end{barticle}
\endbibitem

\bibitem{Akbarimoosavi2019}
\begin{barticle}
\bauthor{\bsnm{Akbarimoosavi}, \binits{M.}},
\bauthor{\bsnm{Rohwer}, \binits{E.}},
\bauthor{\bsnm{Rondi}, \binits{A.}},
\bauthor{\bsnm{Hankache}, \binits{J.}},
\bauthor{\bsnm{Geng}, \binits{Y.}},
\bauthor{\bsnm{Decurtins}, \binits{S.}},
\bauthor{\bsnm{Hauser}, \binits{A.}},
\bauthor{\bsnm{Liu}, \binits{S.-X.}},
\bauthor{\bsnm{Feurer}, \binits{T.}},
\bauthor{\bsnm{Cannizzo}, \binits{A.}}:
\batitle{{Tunable Lifetimes of Intramolecular Charge-Separated States in
  Molecular Donor–Acceptor Dyads}}.
\bjtitle{J. Phys. Chem. C}
\bvolume{123}(\bissue{14}),
\bfpage{8500}--\blpage{8511}
(\byear{2019}).
\doiurl{10.1021/acs.jpcc.8b11066}
\end{barticle}
\endbibitem

\bibitem{rondi2015solvation}
\begin{barticle}
\bauthor{\bsnm{Rondi}, \binits{A.}},
\bauthor{\bsnm{Rodriguez}, \binits{Y.}},
\bauthor{\bsnm{Feurer}, \binits{T.}},
\bauthor{\bsnm{Cannizzo}, \binits{A.}}:
\batitle{{Solvation-Driven Charge Transfer and Localization in Metal
  Complexes}}.
\bjtitle{Acc. Chem. Res.}
\bvolume{48}(\bissue{5}),
\bfpage{1432}--\blpage{1440}
(\byear{2015}).
\doiurl{10.1021/ar5003939}
\end{barticle}
\endbibitem

\bibitem{Aubret2019}
\begin{barticle}
\bauthor{\bsnm{Aubret}, \binits{A.}},
\bauthor{\bsnm{Orrit}, \binits{M.}},
\bauthor{\bsnm{Kulzer}, \binits{F.}}:
\batitle{{Understanding Local-Field Correction Factors in the Framework of the
  Onsager-B{\"{o}}ttcher Model}}.
\bjtitle{ChemPhysChem}
\bvolume{20}(\bissue{3}),
\bfpage{345}--\blpage{355}
(\byear{2019}).
\doiurl{10.1002/cphc.201800923}
\end{barticle}
\endbibitem

\bibitem{Stanley2001}
\begin{barticle}
\bauthor{\bsnm{Stanley}, \binits{R.J.}},
\bauthor{\bsnm{Siddiqui}, \binits{M.S.}}:
\batitle{{A Stark Spectroscopic Study of N(3)-Methyl,
  N(10)-Isobutyl-7,8-Dimethylisoalloxazine in Nonpolar Low-Temperature Glasses:
  Experiment and Comparison with Calculations}}.
\bjtitle{J. Phys. Chem. A}
\bvolume{105}(\bissue{49}),
\bfpage{11001}--\blpage{11008}
(\byear{2001}).
\doiurl{10.1021/jp011971j}
\end{barticle}
\endbibitem

\bibitem{Premvardhan1998}
\begin{barticle}
\bauthor{\bsnm{Premvardhan}, \binits{L.}},
\bauthor{\bsnm{Peteanu}, \binits{L.}}:
\batitle{{Electroabsorption measurements and ab initio calculations of the
  dipolar properties of 2-(2'-hydroxyphenyl)-benzothiazole and -benzoxazole:
  two photostabilizers that undergo excited-state proton transfer}}.
\bjtitle{Chem. Phys. Lett.}
\bvolume{296}(\bissue{5}),
\bfpage{521}--\blpage{529}
(\byear{1998}).
\doiurl{10.1016/S0009-2614(98)01048-3}
\end{barticle}
\endbibitem

\bibitem{Bendikov2004}
\begin{barticle}
\bauthor{\bsnm{Bendikov}, \binits{M.}},
\bauthor{\bsnm{Wudl}, \binits{F.}},
\bauthor{\bsnm{Perepichka}, \binits{D.F.}}:
\batitle{{Tetrathiafulvalenes, Oligoacenenes, and Their Buckminsterfullerene
  Derivatives: The Brick and Mortar of Organic Electronics}}.
\bjtitle{Chem. Rev.}
\bvolume{104}(\bissue{11}),
\bfpage{4891}--\blpage{4946}
(\byear{2004}).
\doiurl{10.1021/cr030666m}
\end{barticle}
\endbibitem

\bibitem{Martin2013}
\begin{barticle}
\bauthor{\bsnm{Mart{\'{i}}n}, \binits{N.}}:
\batitle{{Tetrathiafulvalene: the advent of organic metals}}.
\bjtitle{Chem. Commun.}
\bvolume{49}(\bissue{63}),
\bfpage{7025}--\blpage{7027}
(\byear{2013}).
\doiurl{10.1039/C3CC00240C}
\end{barticle}
\endbibitem

\bibitem{Wu2009}
\begin{barticle}
\bauthor{\bsnm{Wu}, \binits{J.}},
\bauthor{\bsnm{Dupont}, \binits{N.}},
\bauthor{\bsnm{Liu}, \binits{S.-X.}},
\bauthor{\bsnm{Neels}, \binits{A.}},
\bauthor{\bsnm{Hauser}, \binits{A.}},
\bauthor{\bsnm{Decurtins}, \binits{S.}}:
\batitle{{Imidazole-annulated tetrathiafulvalenes exhibiting pH-tuneable
  intramolecular charge transfer and redox properties.}}
\bjtitle{Chem. Asian J.}
\bvolume{4}(\bissue{3}),
\bfpage{392}--\blpage{399}
(\byear{2009}).
\doiurl{10.1002/asia.200800322}
\end{barticle}
\endbibitem

\bibitem{Bergkamp2015}
\begin{barticle}
\bauthor{\bsnm{Bergkamp}, \binits{J.J.}},
\bauthor{\bsnm{Decurtins}, \binits{S.}},
\bauthor{\bsnm{Liu}, \binits{S.-X.}}:
\batitle{{Current advances in fused tetrathiafulvalene donor–acceptor
  systems}}.
\bjtitle{Chem. Soc. Rev.}
\bvolume{44}(\bissue{4}),
\bfpage{863}--\blpage{874}
(\byear{2015}).
\doiurl{10.1039/C4CS00255E}
\end{barticle}
\endbibitem

\bibitem{Segura2001}
\begin{barticle}
\bauthor{\bsnm{Segura}, \binits{J.L.}},
\bauthor{\bsnm{Mart{\'{i}}n}, \binits{N.}}:
\batitle{{New Concepts in Tetrathiafulvalene Chemistry.}}
\bjtitle{Angew. Chem. Int. Ed. Engl.}
\bvolume{40}(\bissue{8}),
\bfpage{1372}--\blpage{1409}
(\byear{2001}).
\doiurl{10.1002/1521-3773(20010417)40:8<1372::aid-anie1372>3.0.co;2-i}
\end{barticle}
\endbibitem

\bibitem{JustinThomas2004}
\begin{barticle}
\bauthor{\bsnm{{Justin Thomas}}, \binits{K.R.}},
\bauthor{\bsnm{Lin}, \binits{J.T.}},
\bauthor{\bsnm{Velusamy}, \binits{M.}},
\bauthor{\bsnm{Tao}, \binits{Y.T.}},
\bauthor{\bsnm{Chuen}, \binits{C.H.}}:
\batitle{{Color Tuning in Benzo[1,2,5]thiadiazole-Based Small Molecules by
  Amino Conjugation/Deconjugation: Bright Red-Light-Emitting Diodes}}.
\bjtitle{Adv. Funct. Mater.}
\bvolume{14}(\bissue{1}),
\bfpage{83}--\blpage{90}
(\byear{2004}).
\doiurl{10.1002/adfm.200304486}
\end{barticle}
\endbibitem

\bibitem{Wu2013}
\begin{barticle}
\bauthor{\bsnm{Wu}, \binits{Y.}},
\bauthor{\bsnm{Zhu}, \binits{W.}}:
\batitle{{Organic sensitizers from D–$\pi$–A to D–A–$\pi$–A: effect
  of the internal electron-withdrawing units on molecular absorption, energy
  levels and photovoltaic performances}}.
\bjtitle{Chem. Soc. Rev.}
\bvolume{42}(\bissue{5}),
\bfpage{2039}--\blpage{2058}
(\byear{2013}).
\doiurl{10.1039/C2CS35346F}
\end{barticle}
\endbibitem

\bibitem{Belton2013}
\begin{barticle}
\bauthor{\bsnm{Belton}, \binits{C.R.}},
\bauthor{\bsnm{Kanibolotsky}, \binits{A.L.}},
\bauthor{\bsnm{Kirkpatrick}, \binits{J.}},
\bauthor{\bsnm{Orofino}, \binits{C.}},
\bauthor{\bsnm{Elmasly}, \binits{S.E.T.}},
\bauthor{\bsnm{Stavrinou}, \binits{P.N.}},
\bauthor{\bsnm{Skabara}, \binits{P.J.}},
\bauthor{\bsnm{Bradley}, \binits{D.D.C.}}:
\batitle{{Location, Location, Location - Strategic Positioning of
  2,1,3-Benzothiadiazole Units within Trigonal Quaterfluorene-Truxene
  Star-Shaped Structures}}.
\bjtitle{Adv. Funct. Mater.}
\bvolume{23}(\bissue{22}),
\bfpage{2792}--\blpage{2804}
(\byear{2013}).
\doiurl{10.1002/adfm.201202644}
\end{barticle}
\endbibitem

\bibitem{Pop2013}
\begin{barticle}
\bauthor{\bsnm{Pop}, \binits{F.}},
\bauthor{\bsnm{Riob{\'{e}}}, \binits{F.}},
\bauthor{\bsnm{Seifert}, \binits{S.}},
\bauthor{\bsnm{Cauchy}, \binits{T.}},
\bauthor{\bsnm{Ding}, \binits{J.}},
\bauthor{\bsnm{Dupont}, \binits{N.}},
\bauthor{\bsnm{Hauser}, \binits{A.}},
\bauthor{\bsnm{Koch}, \binits{M.}},
\bauthor{\bsnm{Avarvari}, \binits{N.}}:
\batitle{{Tetrathiafulvalene-1,3,5-triazines as (Multi)Donor–Acceptor Systems
  with Tunable Charge Transfer: Structural, Photophysical, and Theoretical
  Investigations}}.
\bjtitle{Inorg. Chem.}
\bvolume{52}(\bissue{9}),
\bfpage{5023}--\blpage{5034}
(\byear{2013}).
\doiurl{10.1021/ic3027336}
\end{barticle}
\endbibitem

\bibitem{Alemany2015}
\begin{barticle}
\bauthor{\bsnm{Alemany}, \binits{P.}},
\bauthor{\bsnm{Canadell}, \binits{E.}},
\bauthor{\bsnm{Geng}, \binits{Y.}},
\bauthor{\bsnm{Hauser}, \binits{J.}},
\bauthor{\bsnm{Macchi}, \binits{P.}},
\bauthor{\bsnm{Kr{\"{a}}mer}, \binits{K.}},
\bauthor{\bsnm{Decurtins}, \binits{S.}},
\bauthor{\bsnm{Liu}, \binits{S.-X.}}:
\batitle{{Exploring the Electronic Structure of an Organic Semiconductor Based
  on a Compactly Fused Electron Donor–Acceptor Molecule}}.
\bjtitle{ChemPhysChem}
\bvolume{16}(\bissue{7}),
\bfpage{1361}--\blpage{1365}
(\byear{2015}).
\doiurl{10.1002/cphc.201500090}
\end{barticle}
\endbibitem

\bibitem{Geng2014}
\begin{barticle}
\bauthor{\bsnm{Geng}, \binits{Y.}},
\bauthor{\bsnm{Pfattner}, \binits{R.}},
\bauthor{\bsnm{Campos}, \binits{A.}},
\bauthor{\bsnm{Hauser}, \binits{J.}},
\bauthor{\bsnm{Laukhin}, \binits{V.}},
\bauthor{\bsnm{Puigdollers}, \binits{J.}},
\bauthor{\bsnm{Veciana}, \binits{J.}},
\bauthor{\bsnm{Mas-Torrent}, \binits{M.}},
\bauthor{\bsnm{Rovira}, \binits{C.}},
\bauthor{\bsnm{Decurtins}, \binits{S.}},
\bauthor{\bsnm{Liu}, \binits{S.-X.}}:
\batitle{{A Compact Tetrathiafulvalene–Benzothiadiazole Dyad and Its Highly
  Symmetrical Charge-Transfer Salt: Ordered Donor $\pi$-Stacks Closely Bound to
  Their Acceptors}}.
\bjtitle{Chem. – A Eur. J.}
\bvolume{20}(\bissue{23}),
\bfpage{7136}--\blpage{7143}
(\byear{2014}).
\doiurl{10.1002/chem.201304688}
\end{barticle}
\endbibitem

\bibitem{Lambert2018}
\begin{barticle}
\bauthor{\bsnm{Lambert}, \binits{C.J.}},
\bauthor{\bsnm{Liu}, \binits{S.-X.}}:
\batitle{{A Magic Ratio Rule for Beginners: A Chemist's Guide to Quantum
  Interference in Molecules}}.
\bjtitle{Chem. – A Eur. J.}
\bvolume{24}(\bissue{17}),
\bfpage{4193}--\blpage{4201}
(\byear{2018}).
\doiurl{10.1002/chem.201704488}
\end{barticle}
\endbibitem

\bibitem{Geng2015}
\begin{barticle}
\bauthor{\bsnm{Geng}, \binits{Y.}},
\bauthor{\bsnm{Sangtarash}, \binits{S.}},
\bauthor{\bsnm{Huang}, \binits{C.}},
\bauthor{\bsnm{Sadeghi}, \binits{H.}},
\bauthor{\bsnm{Fu}, \binits{Y.}},
\bauthor{\bsnm{Hong}, \binits{W.}},
\bauthor{\bsnm{Wandlowski}, \binits{T.}},
\bauthor{\bsnm{Decurtins}, \binits{S.}},
\bauthor{\bsnm{Lambert}, \binits{C.J.}},
\bauthor{\bsnm{Liu}, \binits{S.-X.}}:
\batitle{{Magic Ratios for Connectivity-Driven Electrical Conductance of
  Graphene-like Molecules}}.
\bjtitle{J. Am. Chem. Soc.}
\bvolume{137}(\bissue{13}),
\bfpage{4469}--\blpage{4476}
(\byear{2015}).
\doiurl{10.1021/jacs.5b00335}
\end{barticle}
\endbibitem

\bibitem{Famili2019}
\begin{barticle}
\bauthor{\bsnm{Famili}, \binits{M.}},
\bauthor{\bsnm{Jia}, \binits{C.}},
\bauthor{\bsnm{Liu}, \binits{X.}},
\bauthor{\bsnm{Wang}, \binits{P.}},
\bauthor{\bsnm{Grace}, \binits{I.M.}},
\bauthor{\bsnm{Guo}, \binits{J.}},
\bauthor{\bsnm{Liu}, \binits{Y.}},
\bauthor{\bsnm{Feng}, \binits{Z.}},
\bauthor{\bsnm{Wang}, \binits{Y.}},
\bauthor{\bsnm{Zhao}, \binits{Z.}},
\bauthor{\bsnm{Decurtins}, \binits{S.}},
\bauthor{\bsnm{H{\"{a}}ner}, \binits{R.}},
\bauthor{\bsnm{Huang}, \binits{Y.}},
\bauthor{\bsnm{Liu}, \binits{S.-X.}},
\bauthor{\bsnm{Lambert}, \binits{C.J.}},
\bauthor{\bsnm{Duan}, \binits{X.}}:
\batitle{{Self-Assembled Molecular-Electronic Films Controlled by Room
  Temperature Quantum Interference}}.
\bjtitle{Chem}
\bvolume{5}(\bissue{2}),
\bfpage{474}--\blpage{484}
(\byear{2019}).
\doiurl{10.1016/j.chempr.2018.12.008}
\end{barticle}
\endbibitem

\bibitem{Geng2015a}
\begin{barticle}
\bauthor{\bsnm{Geng}, \binits{Y.}},
\bauthor{\bsnm{Yi}, \binits{C.}},
\bauthor{\bsnm{Bircher}, \binits{M.P.}},
\bauthor{\bsnm{Decurtins}, \binits{S.}},
\bauthor{\bsnm{Cascella}, \binits{M.}},
\bauthor{\bsnm{Gr{\"{a}}tzel}, \binits{M.}},
\bauthor{\bsnm{Liu}, \binits{S.-X.}}:
\batitle{{Anthanthrene dye-sensitized solar cells: influence of the number of
  anchoring groups and substitution motif}}.
\bjtitle{RSC Adv.}
\bvolume{5}(\bissue{119}),
\bfpage{98643}--\blpage{98652}
(\byear{2015}).
\doiurl{10.1039/C5RA21917E}
\end{barticle}
\endbibitem

\bibitem{Giguere2015}
\begin{barticle}
\bauthor{\bsnm{Gigu{\`{e}}re}, \binits{J.-B.}},
\bauthor{\bsnm{Sariciftci}, \binits{N.S.}},
\bauthor{\bsnm{Morin}, \binits{J.-F.}}:
\batitle{{Polycyclic anthanthrene small molecules: semiconductors for organic
  field-effect transistors and solar cells applications}}.
\bjtitle{J. Mater. Chem. C}
\bvolume{3}(\bissue{3}),
\bfpage{601}--\blpage{606}
(\byear{2015}).
\doiurl{10.1039/C4TC02137A}
\end{barticle}
\endbibitem

\bibitem{Zhang2012a}
\begin{barticle}
\bauthor{\bsnm{Zhang}, \binits{L.}},
\bauthor{\bsnm{Walker}, \binits{B.}},
\bauthor{\bsnm{Liu}, \binits{F.}},
\bauthor{\bsnm{Colella}, \binits{N.S.}},
\bauthor{\bsnm{Mannsfeld}, \binits{S.C.B.}},
\bauthor{\bsnm{Watkins}, \binits{J.J.}},
\bauthor{\bsnm{Nguyen}, \binits{T.-Q.}},
\bauthor{\bsnm{Briseno}, \binits{A.L.}}:
\batitle{{Triisopropylsilylethynyl-functionalized dibenzo[def,mno]chrysene: a
  solution-processed small molecule for bulk heterojunction solar cells}}.
\bjtitle{J. Mater. Chem.}
\bvolume{22}(\bissue{10}),
\bfpage{4266}--\blpage{4268}
(\byear{2012}).
\doiurl{10.1039/C2JM14998B}
\end{barticle}
\endbibitem

\bibitem{Shah2006}
\begin{barticle}
\bauthor{\bsnm{Shah}, \binits{B.K.}},
\bauthor{\bsnm{Neckers}, \binits{D.C.}},
\bauthor{\bsnm{Shi}, \binits{J.}},
\bauthor{\bsnm{Forsythe}, \binits{E.W.}},
\bauthor{\bsnm{Morton}, \binits{D.}}:
\batitle{{Anthanthrene Derivatives as Blue Emitting Materials for Organic
  Light-Emitting Diode Applications}}.
\bjtitle{Chem. Mater.}
\bvolume{18}(\bissue{3}),
\bfpage{603}--\blpage{608}
(\byear{2006}).
\doiurl{10.1021/cm052188x}
\end{barticle}
\endbibitem

\bibitem{Shah2005}
\begin{barticle}
\bauthor{\bsnm{Shah}, \binits{B.K.}},
\bauthor{\bsnm{Neckers}, \binits{D.C.}},
\bauthor{\bsnm{Shi}, \binits{J.}},
\bauthor{\bsnm{Forsythe}, \binits{E.W.}},
\bauthor{\bsnm{Morton}, \binits{D.}}:
\batitle{{Photophysical Properties of Anthanthrene-Based Tunable Blue
  Emitters}}.
\bjtitle{J. Phys. Chem. A}
\bvolume{109}(\bissue{34}),
\bfpage{7677}--\blpage{7681}
(\byear{2005}).
\doiurl{10.1021/jp052337z}
\end{barticle}
\endbibitem

\bibitem{Adamo1999}
\begin{barticle}
\bauthor{\bsnm{Adamo}, \binits{C.}},
\bauthor{\bsnm{Barone}, \binits{V.}}:
\batitle{{Toward reliable density functional methods without adjustable
  parameters: The PBE0 model}}.
\bjtitle{J. Chem. Phys.}
\bvolume{110}(\bissue{13}),
\bfpage{6158}--\blpage{6170}
(\byear{1999}).
\doiurl{10.1063/1.478522}
\end{barticle}
\endbibitem

\bibitem{Becke1993}
\begin{barticle}
\bauthor{\bsnm{Becke}, \binits{A.D.}}:
\batitle{{Density‐functional thermochemistry. III. The role of exact
  exchange}}.
\bjtitle{J. Chem. Phys.}
\bvolume{98}(\bissue{7}),
\bfpage{5648}--\blpage{5652}
(\byear{1993}).
\doiurl{10.1063/1.464913}
\end{barticle}
\endbibitem

\bibitem{Hehre1972}
\begin{barticle}
\bauthor{\bsnm{Hehre}, \binits{W.J.}},
\bauthor{\bsnm{Ditchfield}, \binits{R.}},
\bauthor{\bsnm{Pople}, \binits{J.A.}}:
\batitle{{Self—Consistent Molecular Orbital Methods. XII. Further Extensions
  of Gaussian—Type Basis Sets for Use in Molecular Orbital Studies of Organic
  Molecules}}.
\bjtitle{J. Chem. Phys.}
\bvolume{56}(\bissue{5}),
\bfpage{2257}--\blpage{2261}
(\byear{1972}).
\doiurl{10.1063/1.1677527}
\end{barticle}
\endbibitem

\bibitem{Runge1984}
\begin{barticle}
\bauthor{\bsnm{Runge}, \binits{E.}},
\bauthor{\bsnm{Gross}, \binits{E.K.U.}}:
\batitle{{Density-Functional Theory for Time-Dependent Systems}}.
\bjtitle{Phys. Rev. Lett.}
\bvolume{52}(\bissue{12}),
\bfpage{997}--\blpage{1000}
(\byear{1984}).
\doiurl{10.1103/PhysRevLett.52.997}
\end{barticle}
\endbibitem

\bibitem{Jansik2004}
\begin{barticle}
\bauthor{\bsnm{Jansik}, \binits{B.}},
\bauthor{\bsnm{Jonsson}, \binits{D.}},
\bauthor{\bsnm{Sa{\l}ek}, \binits{P.}},
\bauthor{\bsnm{{\AA}gren}, \binits{H.}}:
\batitle{{Calculations of static and dynamic polarizabilities of excited states
  by means of density functional theory}}.
\bjtitle{J. Chem. Phys.}
\bvolume{121}(\bissue{16}),
\bfpage{7595}--\blpage{7600}
(\byear{2004}).
\doiurl{10.1063/1.1794635}
\end{barticle}
\endbibitem

\bibitem{Aidas2014}
\begin{barticle}
\bauthor{\bsnm{Aidas}, \binits{K.}},
\bauthor{\bsnm{Angeli}, \binits{C.}},
\bauthor{\bsnm{Bak}, \binits{K.L.}},
\bauthor{\bsnm{Bakken}, \binits{V.}},
\bauthor{\bsnm{Bast}, \binits{R.}},
\bauthor{\bsnm{Boman}, \binits{L.}},
\bauthor{\bsnm{Christiansen}, \binits{O.}},
\bauthor{\bsnm{Cimiraglia}, \binits{R.}},
\bauthor{\bsnm{Coriani}, \binits{S.}},
\bauthor{\bsnm{Dahle}, \binits{P.}},
\bauthor{\bsnm{Dalskov}, \binits{E.K.}},
\bauthor{\bsnm{Ekstr{\"{o}}m}, \binits{U.}},
\bauthor{\bsnm{Enevoldsen}, \binits{T.}},
\bauthor{\bsnm{Eriksen}, \binits{J.J.}},
\bauthor{\bsnm{Ettenhuber}, \binits{P.}},
\bauthor{\bsnm{Fern{\'{a}}ndez}, \binits{B.}},
\bauthor{\bsnm{Ferrighi}, \binits{L.}},
\bauthor{\bsnm{Fliegl}, \binits{H.}},
\bauthor{\bsnm{Frediani}, \binits{L.}},
\bauthor{\bsnm{Hald}, \binits{K.}},
\bauthor{\bsnm{Halkier}, \binits{A.}},
\bauthor{\bsnm{H{\"{a}}ttig}, \binits{C.}},
\bauthor{\bsnm{Heiberg}, \binits{H.}},
\bauthor{\bsnm{Helgaker}, \binits{T.}},
\bauthor{\bsnm{Hennum}, \binits{A.C.}},
\bauthor{\bsnm{Hettema}, \binits{H.}},
\bauthor{\bsnm{Hjerten{\ae}s}, \binits{E.}},
\bauthor{\bsnm{H{\o}st}, \binits{S.}},
\bauthor{\bsnm{H{\o}yvik}, \binits{I.-M.}},
\bauthor{\bsnm{Iozzi}, \binits{M.F.}},
\bauthor{\bsnm{Jans{\'{i}}k}, \binits{B.}},
\bauthor{\bsnm{Jensen}, \binits{H.J.A.}},
\bauthor{\bsnm{Jonsson}, \binits{D.}},
\bauthor{\bsnm{J{\o}rgensen}, \binits{P.}},
\bauthor{\bsnm{Kauczor}, \binits{J.}},
\bauthor{\bsnm{Kirpekar}, \binits{S.}},
\bauthor{\bsnm{Kj{\ae}rgaard}, \binits{T.}},
\bauthor{\bsnm{Klopper}, \binits{W.}},
\bauthor{\bsnm{Knecht}, \binits{S.}},
\bauthor{\bsnm{Kobayashi}, \binits{R.}},
\bauthor{\bsnm{Koch}, \binits{H.}},
\bauthor{\bsnm{Kongsted}, \binits{J.}},
\bauthor{\bsnm{Krapp}, \binits{A.}},
\bauthor{\bsnm{Kristensen}, \binits{K.}},
\bauthor{\bsnm{Ligabue}, \binits{A.}},
\bauthor{\bsnm{Lutn{\ae}s}, \binits{O.B.}},
\bauthor{\bsnm{Melo}, \binits{J.I.}},
\bauthor{\bsnm{Mikkelsen}, \binits{K.V.}},
\bauthor{\bsnm{Myhre}, \binits{R.H.}},
\bauthor{\bsnm{Neiss}, \binits{C.}},
\bauthor{\bsnm{Nielsen}, \binits{C.B.}},
\bauthor{\bsnm{Norman}, \binits{P.}},
\bauthor{\bsnm{Olsen}, \binits{J.}},
\bauthor{\bsnm{Olsen}, \binits{J.M.H.}},
\bauthor{\bsnm{Osted}, \binits{A.}},
\bauthor{\bsnm{Packer}, \binits{M.J.}},
\bauthor{\bsnm{Pawlowski}, \binits{F.}},
\bauthor{\bsnm{Pedersen}, \binits{T.B.}},
\bauthor{\bsnm{Provasi}, \binits{P.F.}},
\bauthor{\bsnm{Reine}, \binits{S.}},
\bauthor{\bsnm{Rinkevicius}, \binits{Z.}},
\bauthor{\bsnm{Ruden}, \binits{T.A.}},
\bauthor{\bsnm{Ruud}, \binits{K.}},
\bauthor{\bsnm{Rybkin}, \binits{V.V.}},
\bauthor{\bsnm{Sa{\l}ek}, \binits{P.}},
\bauthor{\bsnm{Samson}, \binits{C.C.M.}},
\bauthor{\bparticle{de} \bsnm{Mer{\'{a}}s}, \binits{A.S.}},
\bauthor{\bsnm{Saue}, \binits{T.}},
\bauthor{\bsnm{Sauer}, \binits{S.P.A.}},
\bauthor{\bsnm{Schimmelpfennig}, \binits{B.}},
\bauthor{\bsnm{Sneskov}, \binits{K.}},
\bauthor{\bsnm{Steindal}, \binits{A.H.}},
\bauthor{\bsnm{Sylvester-Hvid}, \binits{K.O.}},
\bauthor{\bsnm{Taylor}, \binits{P.R.}},
\bauthor{\bsnm{Teale}, \binits{A.M.}},
\bauthor{\bsnm{Tellgren}, \binits{E.I.}},
\bauthor{\bsnm{Tew}, \binits{D.P.}},
\bauthor{\bsnm{Thorvaldsen}, \binits{A.J.}},
\bauthor{\bsnm{Th{\o}gersen}, \binits{L.}},
\bauthor{\bsnm{Vahtras}, \binits{O.}},
\bauthor{\bsnm{Watson}, \binits{M.A.}},
\bauthor{\bsnm{Wilson}, \binits{D.J.D.}},
\bauthor{\bsnm{Ziolkowski}, \binits{M.}},
\bauthor{\bsnm{{\AA}gren}, \binits{H.}}:
\batitle{{The Dalton quantum chemistry program system}}.
\bjtitle{WIREs Comput. Mol. Sci.}
\bvolume{4}(\bissue{3}),
\bfpage{269}--\blpage{284}
(\byear{2014}).
\doiurl{10.1002/wcms.1172}
\end{barticle}
\endbibitem

\bibitem{Hebling2002}
\begin{barticle}
\bauthor{\bsnm{Hebling}, \binits{J.}},
\bauthor{\bsnm{Alm{\'{a}}si}, \binits{G.}},
\bauthor{\bsnm{Kozma}, \binits{I.Z.}},
\bauthor{\bsnm{Kuhl}, \binits{J.}}:
\batitle{{Velocity matching by pulse front tilting for large-area THz-pulse
  generation}}.
\bjtitle{Opt. Express}
\bvolume{10}(\bissue{21}),
\bfpage{1161}--\blpage{1166}
(\byear{2002}).
\doiurl{10.1364/OE.10.001161}
\end{barticle}
\endbibitem

\bibitem{Fulop2014}
\begin{barticle}
\bauthor{\bsnm{F{\"{u}}l{\"{o}}p}, \binits{J.A.}},
\bauthor{\bsnm{Ollmann}, \binits{Z.}},
\bauthor{\bsnm{Lombosi}, \binits{C.}},
\bauthor{\bsnm{Skrobol}, \binits{C.}},
\bauthor{\bsnm{Klingebiel}, \binits{S.}},
\bauthor{\bsnm{P{\'{a}}lfalvi}, \binits{L.}},
\bauthor{\bsnm{Krausz}, \binits{F.}},
\bauthor{\bsnm{Karsch}, \binits{S.}},
\bauthor{\bsnm{Hebling}, \binits{J.}}:
\batitle{{Efficient generation of THz pulses with 0.4 mJ energy}}.
\bjtitle{Opt. Express}
\bvolume{22}(\bissue{17}),
\bfpage{20155}--\blpage{20163}
(\byear{2014}).
\doiurl{10.1364/OE.22.020155}
\end{barticle}
\endbibitem

\bibitem{Sajadi2017}
\begin{barticle}
\bauthor{\bsnm{Sajadi}, \binits{M.}},
\bauthor{\bsnm{Wolf}, \binits{M.}},
\bauthor{\bsnm{Kampfrath}, \binits{T.}}:
\batitle{{Transient birefringence of liquids induced by terahertz
  electric-field torque on permanent molecular dipoles}}.
\bjtitle{Nat. Commun.}
\bvolume{8}(\bissue{1}),
\bfpage{14963}
(\byear{2017}).
\doiurl{10.1038/ncomms14963}
\end{barticle}
\endbibitem

\bibitem{Wu1995}
\begin{barticle}
\bauthor{\bsnm{Wu}, \binits{Q.}},
\bauthor{\bsnm{Zhang}, \binits{X.-C.}}:
\batitle{{Free-space electro-optic sampling of terahertz beams}}.
\bjtitle{Appl. Phys. Lett.}
\bvolume{67}(\bissue{24}),
\bfpage{3523}--\blpage{3525}
(\byear{1995}).
\doiurl{10.1063/1.114909}
\end{barticle}
\endbibitem

\bibitem{Nahata1996}
\begin{barticle}
\bauthor{\bsnm{Nahata}, \binits{A.}},
\bauthor{\bsnm{Weling}, \binits{A.S.}},
\bauthor{\bsnm{Heinz}, \binits{T.F.}}:
\batitle{{A wideband coherent terahertz spectroscopy system using optical
  rectification and electro‐optic sampling}}.
\bjtitle{Appl. Phys. Lett.}
\bvolume{69}(\bissue{16}),
\bfpage{2321}--\blpage{2323}
(\byear{1996}).
\doiurl{10.1063/1.117511}
\end{barticle}
\endbibitem

\bibitem{Brunner2014}
\begin{barticle}
\bauthor{\bsnm{Brunner}, \binits{F.D.J.}},
\bauthor{\bsnm{Johnson}, \binits{J.A.}},
\bauthor{\bsnm{Gr{\"{u}}bel}, \binits{S.}},
\bauthor{\bsnm{Ferrer}, \binits{A.}},
\bauthor{\bsnm{Johnson}, \binits{S.L.}},
\bauthor{\bsnm{Feurer}, \binits{T.}}:
\batitle{{Distortion-free enhancement of terahertz signals measured by
  electro-optic sampling. I. Theory}}.
\bjtitle{J. Opt. Soc. Am. B}
\bvolume{31}(\bissue{4}),
\bfpage{904}--\blpage{910}
(\byear{2014}).
\doiurl{10.1364/JOSAB.31.000904}
\end{barticle}
\endbibitem

\bibitem{Isnardi1922}
\begin{barticle}
\bauthor{\bsnm{Isnardi}, \binits{H.}}:
\batitle{{Die Dielektrizit{\"{a}}tskonstante von Fl{\"{u}}ssigkeiten in ihrer
  Temperaturabh{\"{a}}ngigkeit}}.
\bjtitle{Zeitschrift f{\"{u}}r Phys.}
\bvolume{9}(\bissue{1}),
\bfpage{153}--\blpage{179}
(\byear{1922}).
\doiurl{10.1007/BF01326964}
\end{barticle}
\endbibitem

\bibitem{Ro/nne2000}
\begin{barticle}
\bauthor{\bsnm{R\o{}nne}, \binits{C.}},
\bauthor{\bsnm{Jensby}, \binits{K.}},
\bauthor{\bsnm{Loughnane}, \binits{B.J.}},
\bauthor{\bsnm{Fourkas}, \binits{J.}},
\bauthor{\bsnm{Nielsen}, \binits{O.F.}},
\bauthor{\bsnm{Keiding}, \binits{S.R.}}:
\batitle{{Temperature dependence of the dielectric function of C6H6(l) and
  C6H5CH3(l) measured with THz spectroscopy}}.
\bjtitle{J. Chem. Phys.}
\bvolume{113}(\bissue{9}),
\bfpage{3749}--\blpage{3756}
(\byear{2000}).
\doiurl{10.1063/1.1287737}
\end{barticle}
\endbibitem

\end{thebibliography}
\end{document}